\documentclass[aps,prd,preprint,superscriptaddress,showpacs,showkeys,nofootinbib]{revtex4-1}
\pdfoutput=1

\usepackage[intlimits]{amsmath}
\usepackage{amssymb}
\usepackage{dcolumn}
\usepackage{bm}
\usepackage{feynmp}                                     
\usepackage{hyperref,graphicx}           
\usepackage{color}

\begin{document}

\title{Colorless Top Partners, a 125 GeV Higgs,\\ and the Limits on 
Naturalness}

\author{Gustavo Burdman}
\affiliation{Instituto de F\'{i}sica, Universidade de S\~{a}o Paulo, S\~{a}o Paulo, S\~{a}o Paulo 05508-900, Brazil}
\author{Zackaria Chacko}
\affiliation{Maryland Center for Fundamental Physics, Department of Physics,\\ University of Maryland, College Park, MD 20742-4111 USA}
\author{Roni Harnik}
\affiliation{Fermilab, P.O. Box 500, Batavia, IL 60510, USA}
\author{Leonardo de Lima}
\affiliation{Instituto de F\'{i}sica Te\'{o}rica, Universidade Estadual Paulista, S\~{a}o Paulo, S\~{a}o Paulo 01140-070, Brazil}
\author{Christopher B. Verhaaren}
\affiliation{Maryland Center for Fundamental Physics, Department of Physics,\\ University of Maryland, College Park, MD 20742-4111 USA}

\date{\today}
\begin{abstract}

Theories of physics beyond the Standard Model that address 
the hierarchy problem generally involve top partners, new particles that cancel 
the quadratic divergences associated with the Yukawa coupling of the 
Higgs to the top quark. With extensions of the Standard Model that 
involve new colored particles coming under strain from collider 
searches, scenarios in which the top partners carry no charge under the 
strong interactions have become increasingly compelling. Although 
elusive for direct searches, these theories predict modified couplings 
of the Higgs boson to the Standard Model particles. This results in 
corrections to the Higgs production and decay rates that can be detected 
at the Large Hadron Collider (LHC) provided the top partners are 
sufficiently light, and the theory correspondingly natural. In this 
paper we consider three theories that address the little hierarchy 
problem and involve colorless top partners, specifically the Mirror Twin 
Higgs, Folded Supersymmetry, and the Quirky Little Higgs. For each model 
we investigate the current and future bounds on the top partners, and 
the corresponding limits on naturalness, that can be obtained from the 
Higgs program at the LHC. We conclude that the LHC will not be able to 
strongly disfavor naturalness, with mild tuning at the level of about 
one part in ten remaining allowed even with 3000 fb$^{-1}$ of data at 14 
TeV.

\end{abstract}


\begin{flushright}
FERMILAB-PUB-14-411-T\\
\end{flushright}


\maketitle
\section{Introduction --  Uncolored Top Partners\label{sec:intro}}

\vspace{0.5cm}

\noindent  {\tt ``It's a bit like spotting a familiar face from afar. Sometimes you need closer inspection to find out whether it's really your best friend, or actually your best friend's twin.''}\\

\hspace*{10cm} \emph{-- Rolf Heuer, July 2012}
\begin{center} ------------------ \end{center}
\vspace{0.4cm}



The discovery of a 125 GeV Higgs boson at the LHC~\cite{Aad:2012tfa,Chatrchyan:2012ufa} seems to complete the 
Standard Model (SM) of particle physics. With the inclusion of the 
Higgs, the SM is a perfectly well-behaved theory up to energies that are 
exponentially higher than the Higgs mass. This extrapolation, without 
the inclusion of new physics, presents a theoretical problem because 
achieving the observed hierarchy between the electroweak scale and the 
Planck scale requires exquisite fine-tuning. This tuning is required 
because quadratically divergent radiative corrections to the Higgs mass 
parameter need to be canceled by a large bare mass. One of the most 
important questions in high energy physics today is whether such a 
tuning indeed exists in nature or whether the electroweak scale is set 
by a mechanism that does not require a large cancellation. This is the 
question of Higgs naturalness, or the hierarchy problem.

An attractive dynamical solution to the naturalness problem is to posit 
a new symmetry which protects the Higgs against large radiative 
corrections. Invoking such a symmetry implies the existence of particles 
beyond the SM which constitute the ``symmetry partners'' of known SM fields. Considered from the bottom up, the hierarchy problem is dominated by the one loop diagram involving the top quark. Any model that addresses the hierarchy problem must therefore include symmetry partners for the top quark, the ``top partners." To avoid significant residual tuning, the top partners are expected to have masses at or below the TeV scale. Well known examples of top partners 
include the scalar stops in supersymmetric models~(for a review see 
\cite{Martin:1997ns}) and vectorlike fermionic top-primes in little 
Higgs models~\cite{ArkaniHamed:2001nc,ArkaniHamed:2002qx,ArkaniHamed:2002qy,Schmaltz:2004de}~(for a review see~\cite{Schmaltz:2005ky}). In these 
examples the new symmetry that is protecting the Higgs commutes with the 
SM gauge symmetries, and so the top partners have identical quantum 
numbers to those of the top. In particular, the top partners in these 
models are charged under the SM color group. This fact, in combination 
with the expectation that these particles lie at or below the TeV scale, 
leads to the conclusion that top partners should be produced at the LHC 
with high rates.

Searches for colored top partners, both scalar and fermionic, have so 
far yielded only stringent limits~(see e.g.~\cite{Chatrchyan:2013xna,CMS:2014wsa,Aad:2014kra,Aad:2014bva}). Broadly speaking, their masses are constrained to lie above around 700 GeV. This limit is by no means model independent. Indeed, top partners could be hiding, for example, in R-parity violating supersymmetric models, if the spectrum is squeezed~\cite{Martin:2007gf}, or in more elaborate constructions~\cite{Fan:2011yu}. As the LHC experiments improve their 
analyses, the expectation is that most of these holes in the search for 
natural models will be covered.

As models of new physics in which the top partners are colored come 
under increasing strain from LHC searches, theories in which the top 
partners are \emph{not} charged under the strong interactions appear 
ever more compelling. Colorless top partners arise in scenarios where 
the symmetry that protects the Higgs mass does not commute with the SM 
SU(3) color group~\cite{Twnhggs2006, fldsusy2007, Poland:2008ev, 
Quirkyhggs2009}. Instead, the action of the symmetry is to interchange 
SM color with a hidden color group, labeled QCD$'$.

In such a framework, the phenomenology associated with the top partners 
is strikingly different. In particular, since the production cross 
sections for uncolored top partners are many orders of magnitude smaller 
than in the colored case, this allows a simple understanding of why 
these particles have so far escaped discovery.

The most striking possibility along these lines is the Mirror Twin Higgs 
model, where the Higgs is protected by the discrete $Z_2$ subgroup of a 
larger global symmetry~\cite{Twnhggs2006} (see 
also~\cite{Barbieri:2005ri, Chacko:2005vw, Chang:2006ra, Craig:2013fga,Craig:2014aea}). The matter 
content of the theory is simply the SM, and an additional mirror, or 
twin, copy of the SM. In such a scenario, \emph{all of the new 
particles which ensure Higgs naturalness up to scales of order 5-10 TeV 
are singlets under the SM}. The only way to produce new particles at the 
LHC, or to see their effects, is through the Higgs boson itself. In this 
case, the effects of new physics might only appear in precision Higgs 
measurements. While more exotic signals, such as the displaced vertices 
characteristic of hidden valleys~\cite{Strassler:2006im}, are certainly 
possible in specific realizations of this scenario, they are by no means 
guaranteed. It is therefore important to study the Higgs phenomenology of this framework in detail.

In other scenarios, the top partners, while remaining uncolored, carry 
electroweak (EW) charges in addition to QCD$'$. Examples of such 
theories include Folded Supersymmetry~\cite{fldsusy2007} and the Quirky 
Little Higgs~\cite{Quirkyhggs2009}. In such a scenario the top partners 
may be directly produced through the weak interactions. In these 
theories there are no fermions or scalars with masses below the scale 
where QCD$'$ gets strong. Therefore the top partners, once produced, 
exhibit quirky dynamics~\cite{Kang:2008ea}, which leads them to lose 
energy to radiation, followed by pair annihilation~\cite{Kang:2008ea, 
Burdman:2008ek, Harnik:2008ax}. As a consequence of the low EW 
production rates and the exotic phenomenology, discovering these states 
directly is challenging and may require a large LHC data set. Therefore, 
in such scenarios it is also important to study the effects of such 
models on Higgs production and decay rates, since this may lead to 
stronger limits. A study of the Higgs physics would also be important in 
establishing that the quirks, once discovered, are involved in 
addressing the hierarchy problem.

A different category of models is one in which the top partners carry 
electroweak charges, but the gauge symmetry corresponding to QCD$'$ is 
broken and is not present at low energies. This is the case in the Dark 
Top model~\cite{Poland:2008ev}, which has the interesting feature that 
the top partners also constitute the observed dark matter.

In this work we consider some specific theories where the top partners 
are colorless, and study the phenomenology associated with the Higgs. In 
what follows we consider in turn three models: the Mirror Twin Higgs in 
Section \ref{sec:twnhggs}, Folded Supersymmetry in Section 
\ref{sec:fldsusy}, and the Quirky Little Higgs in Section 
\ref{sec:qlhiggs}. For each case we obtain expressions for the Higgs 
production cross section in various channels, and the branching ratios 
into various final states. We use this to determine the current and future 
bounds on the top partners, and the corresponding limits on naturalness, 
that can be obtained from the Higgs program at the LHC.

\section{Mirror Twin Higgs\label{sec:twnhggs}}

\subsection{The Model and Cancellation Mechanism\label{ssec:twinmodel}}

The Mirror Twin Higgs (MTH) model assumes a mirror copy of the complete 
SM, called the twin sector, along with a $Z_2$ symmetry that exchanges 
each SM particle with the corresponding twin partner. In addition, the 
Higgs sector of the theory is assumed to respect an approximate global 
symmetry, which may be taken to be either SU(4)$\times$U(1) or O(8). 
This global symmetry is not exact, but is explicitly violated by the SM 
Yukawa couplings, and also by the SM electroweak gauge interactions. In 
particular, a subgroup of this global symmetry is gauged, and contains 
the SU(2)$\times$U(1) electroweak interactions of the SM, and of the 
twin sector. The SM Higgs doublet emerges as a light 
pseudo-Nambu-Goldstone boson (pNGB) when the global symmetry is 
spontaneously broken. In spite of the fact that the gauge and Yukawa 
interactions explicitly violate the global symmetry, the discrete $Z_2$ 
symmetry ensures the absence of quadratically divergent contributions to 
the Higgs mass to one loop order.

The next step is to understand the cancellation of the quadratic 
divergences in this model. We first consider the case where the breaking 
of the global symmetry, which for concreteness we take to be SU(4)$\times$U(1), is realized by a weakly coupled Higgs sector. The SU(2)$\times$SU(2)$\times$U(1) subgroup of SU(4) and the additional U(1) 
are gauged giving rise to the electroweak interactions in the SM and 
twin sectors. We use the labels $A$ and $B$ to denote the SM and twin 
sectors respectively. Then, under the action of the discrete $Z_2$ 
symmetry, the labels $A$ and $B$ are interchanged, $A \leftrightarrow 
B$. In this notation, $H_A$ represents the SM Higgs doublet and $H_B$ 
the twin doublet. The field $H$, defined as
 \begin{equation}
H=\left(\begin{array}{c}
H_A\\
H_B
\end{array} \right) \; ,
 \end{equation}
 is chosen to transform as the fundamental representation under the 
global SU(4) symmetry. The SU(4) invariant potential for $H$ takes the
form
 \begin{equation}
m^2 H^{\dagger}H + \lambda (H^{\dagger} H)^2
 \end{equation}
 If the parameter $m^2$ is negative, the SU(4)$\times$U(1) symmetry is 
spontaneously broken to SU(3)$\times$U(1) and there are 7 massless 
NGBs in the spectrum. Depending on the alignment of the vacuum expectation value (VEV), several of these NGBs will be eaten. If, however, the VEV of $H$ 
lies along $H_B$, the SM Higgs doublet $H_A$ will remain massless.

The gauge and Yukawa interactions give rise to radiative corrections 
that violate the global symmetry and generate a mass for $H_A$. We 
focus on the top Yukawa coupling, which takes the form
  \begin{equation}
 \lambda_A H_A q_A t_A + \lambda_B H_B q_B t_B \,.
\label{eq:twntopsect}
 \end{equation}
 These interactions generate quadratically divergent corrections to the 
Higgs potential at one loop order. The corrections take the form
 \begin{equation}
\Delta V = \frac{3}{8 \pi^2} \Lambda^2 \left(
\lambda_A^2 H_A^{\dag} H_A + \lambda_B^2 H_B^{\dag} H_B \right) \; ,
 \end{equation} 
 where $\Lambda$ is the ultraviolet (UV) cutoff. The $Z_2$ symmetry, however, ensures 
$\lambda_A = \lambda_B \equiv \lambda$ so that
 \begin{equation}
\Delta V = 
\frac{3 \lambda^2}{8 \pi^2} \Lambda^2 
\left(H_A^{\dag}H_A + H_B^{\dag}H_B \right)
= \frac{3 \lambda^2}{8 \pi^2} \Lambda^2 H^{\dagger} H \; .
 \end{equation}
 Thus, this contribution respects the global symmetry and so 
cannot contribute to the mass of the NGBs. The leading contributions to 
the SM Higgs potential therefore arise from terms which are only 
logarithmically divergent. Consequently, there are no quadratically 
divergent contributions to the Higgs mass at one loop order.

\begin{figure}[ht]
  \vspace{5mm}
\begin{tabular}{ccc}
    \begin{tabular}{c}
\begin{fmffile}{twnhiggs1}
\begin{fmfgraph*}(120,70)
\fmfpen{1.0}
\fmfleft{p1,i1,p2} \fmfright{p3,o1,p4}
\fmfv{l=$\bm{h}$}{i1}\fmfv{l=$\bm{h}$}{o1}
\fmf{dashes}{i1,v1}\fmf{dashes}{v2,o1} 
\fmf{fermion,right=1,tension=0.3,l.side=left,foreground=green}{v1,v2,v1}
\fmfv{decor.shape=circle,decor.filled=full,decor.size=1.5thick,l=$\lambda_t$,l.a=115,l.d=2}{v1} \fmfv{decor.shape=circle,decor.filled=full,decor.size=1.5thick,l=$\lambda_t$,l.a=65,l.d=2}{v2}
\fmfv{l=$q_A$,l.a=0,l.d=40}{p1}
\fmfv{l=$t_A$,l.a=0,l.d=40}{p2}
\end{fmfgraph*}
\end{fmffile}
\end{tabular}
\begin{tabular}{c}
\hspace{5mm}$+$ \hspace{5mm}
\end{tabular} 
\begin{tabular}{c}
\begin{fmffile}{twnhiggs2}
\begin{fmfgraph*}(110,80)
\fmfpen{1.0}
\fmfleft{p1,i1,i2,p3} 
\fmfright{p2,o1,o2,p4}
\fmf{phantom}{p3,v2,p4}
\fmf{dashes}{i1,v1,o1}
\fmfv{l=$\bm{h}$}{i1}\fmfv{l=$\bm{h}$}{o1}
\fmffreeze
\fmf{fermion,right,tension=0.1,l.side=right,foreground=blue}{v2,v1}
\fmf{fermion,right,tension=0.1,l.side=right,foreground=blue}{v1,v2}
\fmfv{decor.shape=circle,decor.filled=full,decor.size=1.5thick, l=$-\lambda_t/(2f)$,l.a=-90,l.d=5}{v1}
\fmfv{decor.shape=cross,l=$\lambda_t f$,l.a=90}{v2}
\fmfv{l=$q_B$,l.d=10,l.a=0}{i2}
\fmfv{l=$t_B$,l.d=10,l.a=180}{o2}
\end{fmfgraph*}
\end{fmffile}
\end{tabular}
\end{tabular}
\caption{\label{fig:twntoploop} Cancellation of quadratic divergences in the Mirror Twin Higgs model. The cancellation holds when the top and its partner are charged under different SU(3)s.}
 \end{figure}
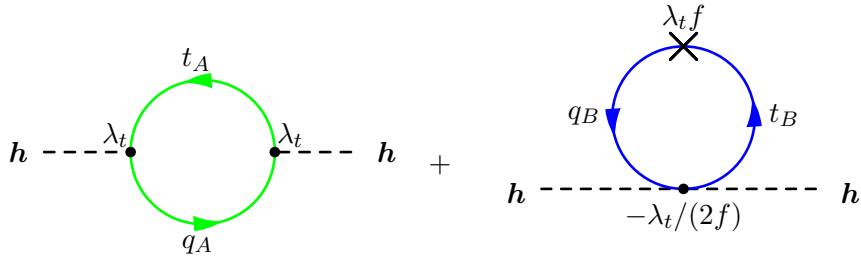

The discussion so far has been restricted to the case when the breaking 
of the global symmetry is realized by a weakly coupled Higgs sector. 
However, the cancellation is in fact independent of the specifics of the UV completion and depends only on the symmetry breaking 
pattern. To see this we consider the low energy effective theory for the 
light degrees of freedom, in which the symmetry is realized 
nonlinearly. We parametrize the pNGB degrees of freedom in terms of 
fields $\Pi^a(x)$ that transform nonlinearly under the broken 
symmetry. For the purpose of writing interactions, it is convenient to 
define an object $H$ which transforms linearly under SU(4)$\times$U(1),
 \begin{equation}
H=\left(\begin{array}{c}
H_A\\
H_B
\end{array} \right) =
\exp\left(\frac{i}{f}\Pi \right)\left(\begin{array}{c}
0\\
0\\
0\\
f
\end{array} \right) \; .
 \end{equation}
 Here $f$ is the symmetry breaking VEV, and $\Pi$ is given, in unitary 
gauge where all the $B$ sector NGBs have been eaten by the corresponding 
vector bosons, by
 \begin{equation}
\Pi=\left(\begin{array}{ccc|c}
0&0&0&h_1\\
0&0&0&h_2\\
0&0&0&0\\ \hline
h_1^{\ast}&h_2^{\ast}&0&0
\end{array} \right).
 \end{equation}
 The discrete $Z_2$ symmetry continues to interchange $H_A$ and $H_B$. 
Expanding out the exponential we obtain
 \begin{equation}
H=\left(\begin{array}{c}
\displaystyle\bm{h}\frac{if}{\sqrt{\bm{h}^{\dag}\bm{h}}}\sin\left( \frac{\sqrt{\bm{h}^{\dag}\bm{h}}}{f} \right)\\
0\\
\displaystyle f\cos\left( \frac{\sqrt{\bm{h}^{\dag}\bm{h}}}{f} \right)
\end{array} \right)
\end{equation}
 where $\bm{h} = (h_1, h_2)^{\text{T}}$ is the Higgs doublet of the SM
 \begin{align}
\displaystyle H_A&= \bm{h}\frac{if}{\sqrt{\bm{h}^{\dag}\bm{h}}}\sin\left( \frac{\sqrt{\bm{h}^{\dag}\bm{h}}}{f} \right)=i \bm{h}+\ldots\,, \\
 H_B&= \left(\begin{array}{c}
0\\
f\cos\left( \frac{\sqrt{\bm{h}^{\dag}\bm{h}}}{f} \right)
\end{array} \right)=\left(\begin{array}{c}
0\\
\displaystyle f-\frac{1}{2f}\bm{h}^{\dag}\bm{h}+\ldots
\end{array} \right).
 \end{align}
 Now consider again the $Z_2$ symmetric top quark sector, 
Eq.~\ref{eq:twntopsect}. To quadratic order in $\bm{h}$ this takes the 
form
 \begin{equation}
i\lambda_t\bm{h} q_At_A+\lambda_t\left(f-\frac{1}{2f}\bm{h}^{\dag}\bm{h} \right) q_Bt_B \; . \label{eq:twntoplag}
 \end{equation} 
From this Lagrangian, we can evaluate the radiative contributions to the Higgs 
mass parameter. The contributing diagrams are shown in Fig. \ref{fig:twntoploop}.

Evaluating these diagrams we find that the quadratic divergence arising 
from the first diagram is exactly canceled by that of the second. The 
first and second diagrams have been colored differently to emphasize 
that the particles running in the two loops carry different SU(3) 
charges. The first loop has the SM top quarks which carry SM color. The 
particles running in the second loop, however, are twin top quarks 
charged under twin color, not SM color.

\subsection{Effects on Higgs Physics\label{ssec:FXHiggstwn}}
In order to understand the implications of this model for Higgs 
production and decays, we first determine the couplings of the 
Higgs to the states in the low energy theory. We choose the
unitary gauge in the visible sector with $h_1=0$ and $h_2=(v+\rho)/\sqrt{2}$ to obtain
 \begin{equation}\begin{array}{cc}
H_A=\left(\begin{array}{c}
0\\
\displaystyle if\sin\left(\frac{v+\rho}{\sqrt{2}f} \right)
\end{array}\right), &H_B=\left(\begin{array}{c}
0\\
\displaystyle f\cos\left(\frac{v+\rho}{\sqrt{2}f} \right)
\end{array}\right).
\end{array}
 \end{equation}
 The couplings of the weak gauge bosons to the Higgs spring from
 \begin{equation}
\left|D_{\mu}^AH_A \right|^2+\left|D_{\mu}^BH_B \right|^2
 \end{equation} 
 where the $D^{A,B}$ denote the covariant derivative employing the $A,B$ 
gauge bosons. Expanding out the kinetic terms we find
 \begin{align}
\frac{1}{2}\partial_{\mu}\rho\partial^{\mu}\rho &+ \left[\frac{f^2g^2}{2}W_{A\mu}^{+}W_A^{\mu-}+\frac{f^2g^2}{4\cos^2\theta_{W}} Z_{A\mu}Z_A^{\mu} \right]\sin^2\left(\frac{v+\rho}{\sqrt{2}f} \right)\nonumber\\
&+\left[\frac{f^2g^2}{2}W_{B\mu}^{+}W_B^{\mu-}+\frac{f^2g^2}{4\cos^2\theta_{W}} Z_{B\mu}Z_B^{\mu} \right]\cos^2\left(\frac{v+\rho}{\sqrt{2}f} \right).\label{eq:twnhgskinetic}
\end{align}

From this we obtain the masses of the $W^\pm$ and $Z$ gauge bosons in 
the visible and twin sectors and their couplings to the Higgs, $\rho$. 
We find that
 \begin{equation}
\begin{array}{cc}
\displaystyle m^2_{W_A}=\frac{f^2g^2}{2}\sin^2\left(\frac{v}{\sqrt{2}f} \right), &\displaystyle m^2_{W_B}=\frac{f^2g^2}{2}\cos^2\left(\frac{v}{\sqrt{2}f} \right) \; .
\end{array}\label{eq:twnhgsbosonmass}
 \end{equation} 
The masses of the $Z$ bosons are related to those of the $W$s by the 
usual factor of $\cos\theta_{W}$. Notice that the VEV of the Higgs in 
the SM, $v_{\text{EW}}=$246 GeV, is related to the parameters $v$ and $f$ 
of the MTH model by the relation
 \begin{equation}
v_{\text{EW}}=\sqrt{2}f\sin\left(\frac{v}{\sqrt{2}f} \right)\equiv\sqrt{2}f\sin\vartheta  \; .
\end{equation}
 From this expression, which defines the angle $\vartheta$, we see that 
$v$ and $v_{\text{EW}}$ become equal in the $v \ll f$, or equivalently 
$\vartheta\ll 1$, limit.

In the absence of any effects that violate the $Z_2$ symmetry, 
minimization of the Higgs potential will reveal that $v_{\text{EW}} = 
f$, so that the state $\rho$ is composed of visible and hidden sector 
states in equal proportions. In order to avoid the experimental limits 
on this scenario, it is desirable to create a hierarchy between these 
scales so that $v_{\text{EW}} < f$. This is most simply realized by a 
soft explicit breaking of the $Z_2$ symmetry. This allows the gauge and 
Yukawa couplings to remain the same across the A and B sectors, so that the 
cancellation of quadratic divergences remains intact.
 
We can expand out \eqref{eq:twnhgskinetic} to obtain the couplings of the Higgs to the electroweak gauge bosons
 \begin{align}
\frac{1}{2}\partial_{\mu}\rho\partial^{\mu}\rho &+\left[m^2_{W_A}W_{A\mu}^{+}W_A^{\mu-}+\frac{m^2_{Z_A}}{2}Z_{A\mu}Z_A^{\mu} \right]\left(1+2\frac{\rho}{v_{\text{EW}}}\cos\vartheta+\cdots \right)\nonumber\\
&+\left[m^2_{W_B}W_{B\mu}^{+}W_B^{\mu-}+\frac{m^2_{Z_B}}{2} Z_{B\mu}Z_B^{\mu} \right]\left(1-2\frac{\rho }{v_{\text{EW}}}\tan\vartheta\sin\vartheta  +\cdots\right).\label{eq:twnhgskinetic2}
\end{align}
 We see that the couplings of $\rho$ to the $W$ and $Z$ differ by a factor of 
$\cos \vartheta$ from the SM prediction.  

We now turn to the top quark sector \eqref{eq:twntopsect}. Expanding this in the unitary gauge we find
\begin{align}
&\lambda_t\left[ifq_At_A\sin\left(\frac{v+\rho}{\sqrt{2}f} \right) +fq_Bt_B\cos\left( \frac{v+\rho}{\sqrt{2}f}\right) \right]\\
=&i\frac{\lambda_t v_{\text{EW}}} {\sqrt{2}}q_At_A\left[1+\frac{\rho}{v_{\text{EW}}}\cos\vartheta \right]+ \lambda_t fq_Bt_B\cos\vartheta\left[ 1-\frac{\rho}{v_{\text{EW}}} \tan\vartheta\sin\vartheta \right] \nonumber
\end{align}
where for simplicity we have not differentiated the components in the SU(2) doublets. We also see that the mass of the top quark's mirror twin partner is
\begin{equation}
m_T=\lambda_t f\cos\vartheta=m_t\cot\vartheta\, .
\end{equation} 

We are also in a position to determine the implications of the MTH model 
for Higgs production and decays. We have seen that the tree level 
couplings of $\rho$ to the visible sector fermions and bosons are simply 
altered by a factor $\cos\vartheta$ relative to the SM. Since the new 
particles in the model carry no SM charges, the radiatively generated 
couplings of the Higgs to gluons and photons are modified relative to 
the SM by exactly the same factor. It follows that all production cross 
sections are modified by the square of this factor,
 \begin{equation}
\sigma(pp\to\rho)=\cos^2\vartheta \; \sigma_{\text{SM}} (pp\to h)
 \end{equation}
where $h$ is the SM Higgs boson. There is a similar relation for decays of the Higgs into $A$ sector 
particles,
 \begin{equation}
\Gamma(\rho\to A_i)=\Gamma^{\text{SM}} (h\to\text{SM}_i) \cos^2\vartheta,
 \end{equation} 
 where the subscript $i$ represents any particle species.
 In addition, $\rho$ will decay into $B$ sector particles that are light 
enough. A factor of $\sin\vartheta$ accompanies couplings of $\rho$ to 
twin sector states, relative to the corresponding SM interactions. We 
define the fraction $\delta$ as
 \begin{equation}
 \delta =\frac{\Gamma(\rho\to B)}{\Gamma^{\text{SM}}(h)\sin^2\vartheta}\; .
 \end{equation} 
 In the limit that the states in the twin sector have the same masses as 
their visible sector partners, $\delta = 1$. Away from this limit, 
$\delta$ is expected to differ from unity due to kinematic effects. The 
total Higgs width in the MTH model is given by
 \begin{equation}
\Gamma(\rho)=\Gamma^{\text{SM}}(h)\left[ \cos^2\vartheta+ \delta\sin^2\vartheta\right].
 \end{equation}

Employing the expressions 
$\Gamma^{\text{SM}}_{\text{BR}}(h\to\text{SM}_i)$ and 
$\Gamma_{\text{BR}}(\rho\to A_i)$ to denote the branching fractions into 
the same particle species $i$ we obtain
 \begin{equation}
\frac{\sigma(pp\to\rho)\Gamma_{\text{BR}}(\rho\to A_i)}{\sigma_{\text{SM}}(pp\to h)\Gamma^{\text{SM}}_{\text{BR}}(h\to\text{SM}_i)}=\frac{\displaystyle\cos^2 \vartheta}{\displaystyle 1+\delta\tan^2 \vartheta}= \frac{1}{\displaystyle\left( 1+\delta\frac{m_t^2}{m_T^2}\right) \left( 1+\frac{m_t^2}{m_T^2}\right)}\, .
 \end{equation}

As explained earlier, in the case when the $Z_2$ symmetry is only softly 
broken, the gauge and Yukawa couplings are the same in the visible and 
twin sectors. This allows us to obtain expressions for the masses of the 
particles in the twin sector, and predict $\delta$. The masses of the 
$B$ sector particles are related to those in the $A$ sector by
 \begin{equation}
m_B = m_A\cot \vartheta
 \end{equation}
 and so for $f\gg v$ the $B$ sector masses are significantly larger that 
those of the $A$ sector. The $B$ sector particles couple to $\rho$ with 
the same coupling as in the SM, but modified by the factor 
$-\sin\vartheta$.

The leading order relation for SM Higgs decays to fermions $f$ is given by
 \begin{equation}
\Gamma(h\to f\overline{f})= \frac{N_c}{16\pi}m_h\lambda_f^2\left(1-4\frac{m_f^2}{m_h^2} \right)^{3/2} \; ,
 \end{equation}
 where $\lambda_f$ is to be evaluated at the Higgs mass. For decays into gauge bosons we use \cite{Djouadi20081}
\begin{equation}
\Gamma(h\to VV^{\ast})=\frac{3 m_h}{32\pi^3}\frac{m_V^4}{v^4_{\text{EW}}} \delta_VR_T\left(\frac{m_V^2}{m_H^2}\right)
\end{equation}
where $\delta_W'=1$, $\delta_Z'=\frac{7}{12}-\frac{10}{9}\sin^2\theta_W +\frac{40}{9}\sin^4\theta_W$, and 
\begin{align}
R_T(x)=&\frac{3(1-8x+20x^2)}{\sqrt{4x-1}}\cos^{-1}\left(\frac{3x-1}{2x^{3/2}}\right)- \frac{1-x}{2x}(2-13x+47x^2)\nonumber\\
&-\frac32(1-6x+4x^2)\ln x\, 
\end{align}
 when the mass of the vector is less than the mass of the Higgs. By 
suitably modifying these expressions, we can obtain the width of the 
Higgs into twin fermions and twin electroweak gauge bosons. The Higgs 
may also decay into twin gluons $g_B$:
 \begin{equation}
\Gamma(\rho\to g_Bg_B)=\frac{\alpha_s^2 m_h^3}{72\pi^3v^2}\left|\frac34 \sum_q A_F\left(\frac{4m_q^2}{m_h^2} \right) \right|^2\, 
 \end{equation}
with $A_F$ defined in \eqref{eq:ferfunc}. The sum is over the twin quarks, but is dominated by the twin top.

We use these formulas in conjunction with the factor of $\sin^2\vartheta$ to determine $\delta$ as a function of $m_t/m_T$:
 \begin{align}
\delta=&\sum_j\Gamma^{\text{SM}}_{\text{BR}}(h\to f_j\overline{f}_j) \left[ \frac{\displaystyle 1-4\frac{m^2_{f_j}}{m_h^2} \frac{m_T^2}{m_t^2}}{\displaystyle 1-4\frac{m^2_{f_j}}{m_h^2}}\right]^{3/2}+\sum_j\Gamma^{\text{SM}}_{\text{BR}}(h\to V_jV^{\ast}_j)\frac{\displaystyle R_T\left(\frac{m_{V_{j}}^2}{m_h^2}\frac{m_T^2}{m_t^2}\right)}{\displaystyle R_T\left( \frac{m_{V_{j}}^2}{m_h^2}\right)}\nonumber\\
&+\Gamma^{\text{SM}}_{\text{BR}}(h\to gg)\frac{\displaystyle \left|A_F\left( \frac{4m_T^2}{m_h^2}\right) \right|^2}{\displaystyle \left|A_F\left( \frac{4m_t^2}{m_h^2}\right) \right|^2}
 \end{align} 
In our analysis, we take into account the decay modes of $\rho$ into the twin sector bottom and charm quarks, and into the tau and muon leptons. We use the Higgs widths reported in~\cite{HiggsProp2013}. 

\begin{figure}[t]
\includegraphics[width=0.6\textwidth]{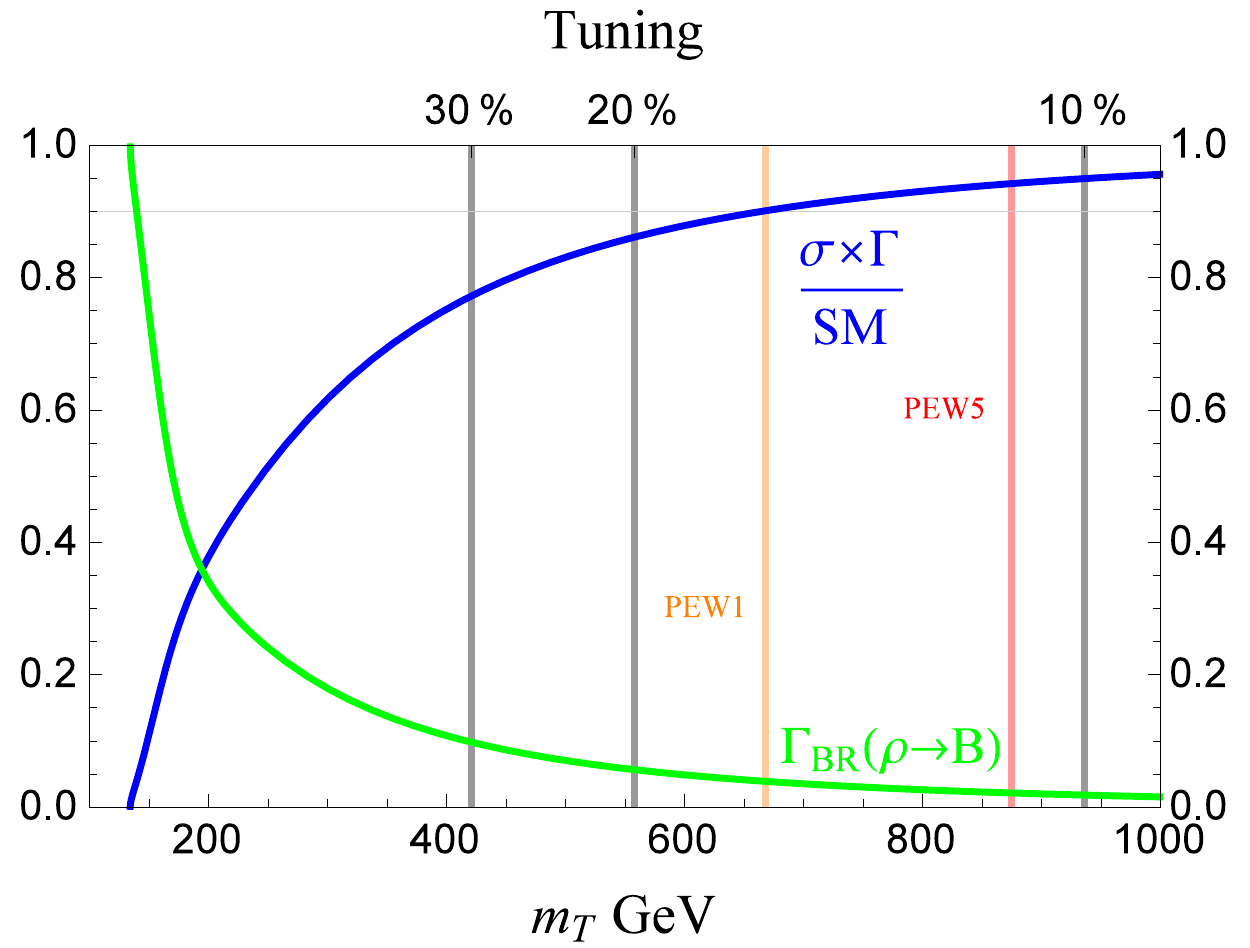}
\caption{\label{Fig:twnhggsratio} In blue, a plot of the rate of Higgs events into SM states normalized to the SM. The green line is the invisible branching ratio of the Higgs into mirror twin particles. The vertical orange and red lines are the 95\% confidence bound from precision electroweak constraints for a 1 and 5 TeV cutoff respectively. }
\end{figure}

Using these results we can determine the rate of Higgs events into any SM state and the  branching fraction into twin sector states. We plot these results in Fig. \ref{Fig:twnhggsratio}. The blue line represents the rate of Higgs events into SM final states in the softly broken MTH model normalized to the SM. The green line denotes the branching fraction of the Higgs into the twin sector particles. A key observation is that the MTH model predicts a relation between the Higgs invisible branching fraction and the modification to standard model rates.

The corrections to the Higgs couplings in the MTH model relative to the 
SM are constrained by precision electroweak measurements. In theories 
where the Higgs emerges as a pNGB, its couplings to the fermions and 
gauge bosons are generally smaller than in the SM. In 
\cite{Falkowski:2013dza} precision electroweak constraints were applied 
to the MCHM4 model~\cite{Agashe:2004rs}, which, like MTH, modifies the 
Higgs couplings to all the vector bosons and fermions by a universal 
factor. Their bound on $\epsilon$, where $\sqrt{1-\epsilon^2}=\cos\vartheta$, also applies to the MTH model in a strongly coupled UV completion, and 
can be translated into a bound on the top partner mass. Their analysis was carried out assuming a cutoff $\Lambda=$3 TeV. In general, however, the leading contributions to the oblique parameters go like
\begin{equation}
\begin{array}{cc}
\displaystyle\alpha T\sim -\epsilon^2\ln\left(\frac{\Lambda}{m_Z} \right), & \displaystyle\alpha S\sim \epsilon^2 \ln\left(\frac{\Lambda}{m_Z} \right),
\end{array}
\end{equation}
where $m_Z$ is the mass of the Z boson. For $\epsilon$ sufficiently small we expect these parameters to dominate the analysis. In that case we may translate the bound on $\epsilon$ at $\Lambda$ to a bound on $\epsilon '$ at $\Lambda '$ by
\begin{equation}
\epsilon^2\ln\left(\frac{\Lambda}{m_Z} \right)= \epsilon^2\left(1+\frac{\ln\left(\frac{\Lambda}{\Lambda '} \right)}{\ln\left(\frac{\Lambda '}{m_Z} \right)} \right)\ln\left(\frac{\Lambda '}{m_Z} \right)\equiv\epsilon '^2\ln\left(\frac{\Lambda '}{m_Z} \right). \label{eq:tuningscaling}
\end{equation}
The $2\sigma$ bound on $\epsilon '$ can be translated into a limit on the top partner mass. In Fig. \ref{Fig:twnhggsratio} we denote bound corresponding to a 1 and 5 TeV cutoff by the vertical orange and red lines respectively. 

Finally, we estimate the tuning $\Delta_m$ of the Higgs mass parameter 
$m^2$ as a function of the top partner mass as a measure of the 
naturalness of the MTH model. We use the formula
 \begin{equation}
\Delta_m=\left|\frac{2\delta m^2}{m_h^2} \right|^{-1}
\end{equation}
 to estimate the tuning. We have denoted the quantum corrections to the 
Higgs mass parameter as $\delta m^2$ and the physical Higgs mass as 
$m_h=$ 125 GeV.

The diagrams in Fig. \ref{fig:twntoploop} lead to 
 \begin{equation}
|\delta m^2|=\frac{3\lambda_t^2 m_T^2}{8\pi^2}\ln\left(\frac{\Lambda^2}{m_T^2}\right) \; ,
 \end{equation}
 up to finite effects. We take the cutoff $\Lambda$ to be 5 TeV. In Fig. 
\ref{Fig:twnhggsratio} we have denoted the top partner masses 
corresponding to 30\%, 20\%, and 10\% tuning.

The results of Fig.~\ref{Fig:twnhggsratio} should be compared to our expectations for the precision at which the LHC will be able to constrain these couplings. Projections for the full high luminosity LHC run (3000 fb$^{-1}$)~\cite{Dawson:2013bba} show that the Higgs invisible branching fraction will be probed down to about 10\%. The precision for the signal strengths in the cleanest Higgs channels, $ZZ$, $WW$, and $\gamma\gamma$, is projected to be around 5\%. The visible signal strengths are thus a stronger constraint on the model and can probe a level of tuning of about 10\% (although combining several channels may improve this sensitivity). The sensitivity at the end of Run~II is only slightly worse. We conclude that models that are tuned at the level of one part in ten  may be able to escape detection at the LHC.

\section{Folded Supersymmetry\label{sec:fldsusy}}

\subsection{The Model and Cancellation Mechanism\label{ssec:foldedmodel}}
Supersymmetry (SUSY) is perhaps the best known solution to the hierarchy 
problem. In supersymmetric theories every known particle is related by 
the symmetry to another particle with a different spin, called its 
superpartner. The gauge quantum numbers of each particle and its 
corresponding superpartner are identical. In supersymmetric extensions 
of the SM, the quadratically divergent contributions to the Higgs mass 
from loops involving the SM particles are canceled by new diagrams 
involving the superpartners.

\begin{figure}[ht]
  \vspace{5mm}
\begin{tabular}{ccc}
    \begin{tabular}{c}
\begin{fmffile}{susytop}
\begin{fmfgraph*}(120,70)
\fmfpen{1.0}
\fmfleft{p1,i1,p2} \fmfright{p3,o1,p4}
\fmfv{l= $H_u$}{i1}\fmfv{l=$H_u$}{o1}
\fmf{dashes}{i1,v1}\fmf{dashes}{v2,o1} 
\fmf{fermion,right=1,tension=0.3,l.side=left,foreground=green}{v1,v2,v1}
\fmfv{decor.shape=circle,decor.filled=full,decor.size=1.5thick,l=$\lambda_t$,l.a=115,l.d=2}{v1} \fmfv{decor.shape=circle,decor.filled=full,decor.size=1.5thick,l=$\lambda_t$,l.a=65,l.d=2}{v2}
\fmfv{l=$q$,l.a=0,l.d=42}{p1}
\fmfv{l=$u$,l.a=0,l.d=42}{p2}
\end{fmfgraph*}
\end{fmffile}
\end{tabular}
\begin{tabular}{c}
\hspace{5mm}$+$ \hspace{5mm}
\end{tabular} 
\begin{tabular}{c}
\begin{fmffile}{susystop}
\begin{fmfgraph*}(110,80)
\fmfpen{1.0}
\fmfleft{p1,i1,i2,p3} 
\fmfright{p2,o1,o2,p4}
\fmf{phantom}{p3,v2,p4}
\fmf{dashes}{i1,v1,o1}
\fmfv{l=$H_u$}{i1}\fmfv{l=$H_u$}{o1}
\fmffreeze
\fmf{double,right,tension=0.1,l.side=right,foreground=blue}{v2,v1}
\fmf{double,right,tension=0.1,l.side=right,foreground=blue}{v1,v2}
\fmfv{decor.shape=circle,decor.filled=full,decor.size=1.5thick, l=$\lambda_t^2$,l.a=-90,l.d=5}{v1}
\fmfv{l=$\widetilde{q},,\widetilde{u}$,l.d=8,l.a=0}{i2} 
\end{fmfgraph*}
\end{fmffile}
\end{tabular}
\end{tabular}
\caption{\label{fig:fsusyloops} Cancellation of quadratic divergences in the Folded SUSY model. This divergence is canceled even if the top and stop transform under different color groups.}
\end{figure}
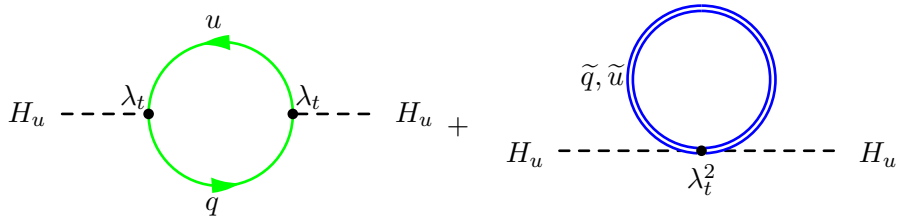

In the case of the top quark, whose left and right components belong to 
the SU(2) doublet $q$ and SU(2) singlet $u$, the corresponding scalar 
partners are the scalar stops, which we label by $\widetilde{q}$ and 
$\widetilde{u}$. Supersymmetric extensions of the SM generally contain 
two Higgs doublets, one labeled $H_u$ which gives mass to the up-type 
quarks and another, labeled $H_d$, which gives mass to the down-type 
quarks and leptons. Both $H_u$ and $H_d$ have fermionic superpartners, 
the Higgsinos. In supersymmetric theories, the one loop quadratically 
divergent contributions to the up-type Higgs mass associated with the 
top Yukawa coupling are canceled by diagrams involving the stops. The 
relevant couplings take the form
 \begin{equation}
(\lambda_t H_u q u+\text{h.c.}) +\lambda_t^2\left|\widetilde{q}H_u \right|^2 +\lambda_t^2\left|\widetilde{u} \right|^2\left|H_u \right|^2 \; .
\label{eq:susytop}
 \end{equation}
These interactions lead to radiative corrections to the up-type Higgs
mass from the diagrams shown in Fig. \ref{fig:fsusyloops}. 

From the form of the interaction in \eqref{eq:susytop}, we see that for 
the cancellation to go through, the left-handed stop $\widetilde{q}$ 
must carry charge under the SU(2) gauge interactions of the SM. At the diagrammatic level, however, the cancellation does not depend on whether the stops transform under SM color.

In Folded Supersymmetric theories the cancellation of the one loop 
quadratic divergences associated with the top Yukawa coupling takes 
place exactly as in the diagrams above, but the top and its scalar 
partners, labeled ``folded stops" or ``F-stops", are charged under 
different color groups. While the fermions transform under the familiar 
SM color group, now labeled SU(3$)_A$, the scalars transform under a 
separate hidden color group, labeled SU(3$)_B$. The electroweak quantum 
numbers of the F-stops are identical to those of the corresponding SM 
fermions. This scenario can be realized in a 5D supersymmetric 
construction, with the extra dimension compactified on $S^1/Z_2$~(see~\cite{Craig:2014fka} for an alternative UV completion). A 
combination of boundary conditions and discrete symmetries ensures that 
the spectrum of light states includes the SM particles and the scalar 
folded superpartners (``F-spartners") that cancel the quadratic 
divergences arising from the couplings of SM fermions to the up- and 
down-type Higgs bosons. The gauginos are projected out by the boundary 
conditions, and are not part of the low energy spectrum. The 
interactions of the top quarks and the F-stops with the up-type Higgs 
have exactly the same form as in \eqref{eq:susytop}, and the 
cancellation of quadratic divergences between the fermion and scalar 
diagrams happens exactly the same way.

\subsection{Effects on Higgs Physics\label{ssec:FXHiggsfsusy}}

In general, the low energy spectrum of Folded Supersymmetry contains two 
Higgs doublets. Our analysis in this section will focus on the limit 
when one of the doublets is much lighter than the other, so that the 
corrections to the Higgs phenomenology primarily arise from the effects 
of the F-stops. In our discussion we follow the conventions of 
Haber\cite{HaberSusy93}. In particular, we take 
$v_{\text{EW}}=\sqrt{v_d^2+v_u^2}$ = 246 GeV where $v_u$ and $v_d$ are 
the VEVs of the up-type and down-type Higgs fields respectively. The 
ratio of the up-type and down-type Higgs VEV is parametrized in terms of 
an angle $\beta$ such that $\tan\beta=v_u/v_d$.

It is well known that in order to obtain a mass of 125 GeV for the light 
Higgs $h^0$ the MSSM is driven into a constrained parameter space with very 
heavy stops, resulting in significant tuning. This issue carries over to 
the folded SUSY construction. One of several possible ways to alleviate 
this constraint is to add another U(1$)_X$ gauge symmetry to the MSSM 
whose $D$-term contribution to the Higgs quartic increases the Higgs 
mass~\cite{Batra:2003nj}.

To be phenomenologically viable, the new gauge field $Z'$ must have a 
mass $m_{Z'}$ not far above the scale of the soft 
masses~\cite{Agashe:2014kda}. This may be realized by giving two heavy 
scalar fields $\phi$ and $\phi^c$ VEVs that break the U(1$)_X$. The 
charge assignments of the SM fields under U(1$)_X$ are chosen to be the 
same as under hypercharge. After integrating out the $\phi$ fields the 
tree level Higgs quartic becomes
 \begin{equation}
\frac18\left[g_{L}^2+g_Y^2+g_X^2\left(1+\frac{m^2_{Z'}}{2m_{\phi}^2} \right)^{-1}\right]\left(|H_u|^2-|H_d|^2 \right)\, ,
\end{equation} 
 where $g_L$, $g_Y$, and $g_X$ are the SU(2$)_L$, U(1$)_Y$, and U(1$)_X$ 
gauge groups. The mass $m_{\phi}$ is the soft mass of $\phi$, which is 
chosen to be equal to that of $\phi^c$ for simplicity.

This method, while not the unique way to raise the Higgs mass, serves to 
illustrate that models of this type may have only moderate tuning from 
the top sector. For concreteness we pick $g_X$ such that the Higgs 
mass, including one loop effects from the top and stops, is 125 GeV. For 
$m_{Z'}=$4 TeV and $m_{\phi}=$5 TeV a perturbative $g_X$ can be chosen 
to give the correct Higgs mass. Additional details of the construction 
are given in Appendix \ref{sec:mssmu1}.

In the limit that only one Higgs doublet is light, its tree level 
couplings to the fermions and gauge bosons are necessarily of the same 
form as in the SM, up to small corrections. Therefore, we need only 
determine the couplings of the Higgs to the F-stops. The stop mixing matrix is
given by 
 \begin{equation}
\left(\begin{array}{cc}
\displaystyle M^2_{\widetilde{Q}}+m_t^2+ m_Z^2\left(\frac12-\frac23 s_W^2 -\frac16\hat{s}^2\right)\cos 2\beta &m_t(A_t-\mu \cot\beta)\\
m_t(A_t-\mu \cot\beta) &\displaystyle M^2_{\widetilde{U}}+m_t^2 +m_Z^2\frac23 \cos 2\beta\left(s^2_W+ \hat{s}^2\right)
\end{array} \right)
 \end{equation}
where $\sin\theta_W\equiv s_W$, $m_t=\lambda_t v_{\text{EW}} 
\sin(\beta)/\sqrt{2}$, and the effective coupling 
\begin{equation}
\hat{s}^2\equiv g_X^2\left(1+\frac{m_{Z'}^2}{2m_{\phi}^2} \right)^{-1} \frac{v^2_{\text{EW}}}{4m_Z^2}\; .
\end{equation}

Although the original incarnation of Folded 
Supersymmetry has $A_t = 0$, in our analysis we allow for the 
possibility that there may be more general constructions that admit 
nonvanishing $A_t$. Then the heavy stop $\widetilde{T}$ and the light 
stop $\widetilde{t}$ can be written as
 \begin{align}
\widetilde{T}&= \cos\alpha_t\widetilde{q}+\sin\alpha_t\widetilde{u}\\
\widetilde{t}&= -\sin\alpha_t\widetilde{q}+\cos\alpha_t\widetilde{u}
 \end{align}
where
 \begin{equation}
\begin{array}{cc}
\displaystyle\cos 2\alpha_t=\frac{M^2_{\widetilde{Q}}-M^2_{\widetilde{U}}+m_Z^2\cos 2\beta\left(\frac12  -\frac43 s_W^2-\frac56 \hat{s}^2\right)} {m^2_{\widetilde{T}}-m^2_{\widetilde{t}}}, &\displaystyle\sin 2\alpha_t=\frac{2m_t(A_t-\mu\cot\beta)}{m^2_{\widetilde{T}}-m^2_{\widetilde{t}}}
\end{array}.\label{eq:stopangle}
 \end{equation}
and
\begin{align}
m^2_{\widetilde{T},\widetilde{t}}=&\frac12\left[M^2_{\widetilde{Q}} +M^2_{\widetilde{U}}+2m^2_t+\frac12 m_Z^2\cos 2\beta\left(1+ \hat{s}^2\right)\right] \nonumber\\
&\pm\frac12\sqrt{ \left[M^2_{\widetilde{Q}}-M^2_{\widetilde{U}}+m_Z^2\cos 2\beta\left(\frac12  -\frac43 s_W^2-\frac56 \hat{s}^2\right) \right]^2+4m_t^2(A_t-\mu\cot\beta)^2}\, .
\end{align}
To ensure that the light stop $\widetilde{t}$ has non-negative mass the 
relation
 \begin{align}
&m_t\left|A_t-\mu\cot\beta \right|\leq\nonumber\\ &\sqrt{\left[M^2_{\widetilde{Q}}+m_t^2+m_Z^2\left(\frac12 -\frac23 s_W^2-\frac16\hat{s}^2 \right) \cos 2\beta\right]\left[M^2_{\widetilde{U}}+m_t^2+m_Z^2\frac23 \cos 2\beta \left(s_W^2+\hat{s}^2\right) \right]}\label{eq:fsusytach}
 \end{align}
must be satisfied.

We can then obtain the couplings of the heavy and light stop mass eigenstates to the light Higgs, $y_{\widetilde{T}}h^0|\widetilde{T}|^2$ and $y_{\widetilde{t}}h^0|\widetilde{t}|^2$. These are given by
 \begin{align}
y_{\widetilde{T}}\equiv \frac{2}{v_{\text{EW}}}&\left\{m_t^2+ m_Z^2\cos 2\beta\left[\frac14+\frac14\hat{s}^2+\left(\frac14-\frac23 s_W^2-\frac{5}{12} \hat{s}^2\right) \cos 2\alpha_t  \right]\right.\nonumber\\
&\left.\phantom{[]}+\frac12 m_t(A_t-\mu\cot\beta)\sin 2\alpha_t \right\},\label{eq:hstopdec}\\
y_{\widetilde{t}}\equiv \frac{2}{v_{\text{EW}}}&\left\{m_t^2+ m_Z^2\cos 2\beta\left[\frac14+\frac14\hat{s}^2 -\left(\frac14-\frac23 s_W^2-\frac{5}{12} \hat{s}^2\right) \cos 2\alpha_t  \right]\right.\nonumber\\
&\left.\phantom{[]}-\frac12 m_t(A_t-\mu\cot\beta)\sin 2\alpha_t \right\}.\label{eq:lstopdec}
 \end{align}

We are now in a position to determine the Higgs phenomenology of this 
model. At tree level, the couplings of the Higgs to the fermions and to 
the $W^\pm$ and $Z$ gauge bosons are the same as in the SM model. 
Furthermore, since the F-stops carry no charge under SM color, the 
couplings of the Higgs to the gluons, which are generated at one loop, 
are also the same as in the SM. It follows that the Higgs production 
cross sections in the gluon fusion, associated production and vector 
boson fusion channels are largely unchanged from the SM predictions.

The Higgs decay widths into SM fermions, gluons and massive gauge bosons 
are also very close to the SM predictions. However, since the F-stops 
do carry electric charges, the rate of Higgs decays to two photons is 
affected. This can be used to constrain the model \cite{Fan:2014txa}. Using the results in Appendix \ref{sec:genexp} we find
 \begin{align}
\Gamma(h^0\to\gamma\gamma)=\frac{\alpha^2m_{h^0}^3}{1024\pi^3}&\left| \frac{2}{v_{\text{EW}}} A_V\left(\frac{4m_W^2}{m_{h^0}^2} \right) +\frac{2}{v_{\text{EW}}} \frac43 A_F\left(\frac{4m_t^2}{m_{h^0}^2} \right)\right.\nonumber\\ 
&\phantom{[]}\left.+ \frac{y_{\widetilde{t}}}{m_{\widetilde{t}}^2}\frac43 A_S\left( \frac{4m_{\widetilde{t}}^2}{m_{h^0}^2}\right)+ \frac{y_{\widetilde{T}}}{m_{\widetilde{T}}^2}\frac43 A_S\left( \frac{4m_{\widetilde{T}}^2}{m_{h^0}^2}\right)\right|^2\label{mess2}
 \end{align} 
where we have employed \eqref{eq:hstopdec} and \eqref{eq:lstopdec} to 
obtain the last two terms.

Having now accounted for all the decay modes we find the corrections to 
the total width are negligible. Therefore, we focus on only the diphoton 
channel. It can be seen from \eqref{eq:hstopdec}, \eqref{eq:lstopdec} 
and \eqref{mess2} that in general the stop loops will contribute with 
the same sign as the top loops and therefore lead to a reduction in the 
diphoton decay rate. If the mixing $A_t$ is increased, however, the 
coupling of the Higgs to the light stop can change sign, leading to an 
enhancement in the rate. We parametrize this difference from the SM 
value by
 \begin{equation}
\delta=\frac{\Gamma(h^0\to\gamma\gamma)-\Gamma^{\text{SM}}(h\to\gamma\gamma)} {\Gamma^{\text{SM}}(h\to\gamma\gamma)}.\label{eq:fldsusydlta}
\end{equation}
Then, neglecting corrections to the overall Higgs width, we have
 \begin{equation}
\frac{\sigma(pp\to h^0)\Gamma_{\text{BR}}(h^0\to \gamma\gamma)}{ \sigma_{\text{SM}}(pp\to h)\Gamma^{\text{SM}}_{\text{BR}}(h\to \gamma\gamma)}=  1+\delta.\label{eq:fldggratio}
 \end{equation}

\begin{figure}[th]
\centering
\includegraphics[width=0.6\textwidth]{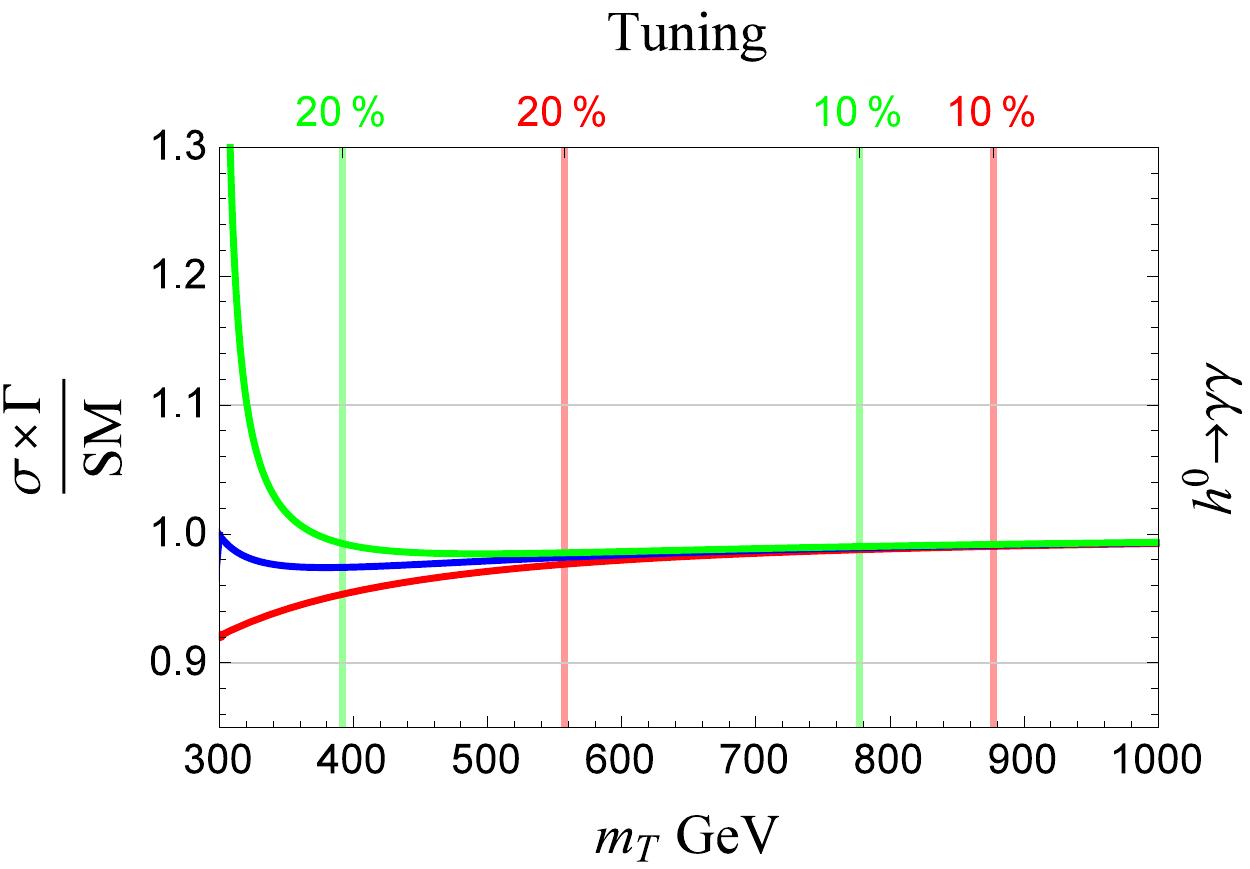}
\caption{\label{fig:fsusyhggratio} Plots of the total Higgs to diphoton rate normalized to SM value as function of the square averaged stop mass $m_T^2$. The  red, blue, and green lines correspond to mixing $A_t-\mu\cot\beta$ equal to 100, 400, and 500 GeV. We have taken the soft masses equal, $\tan\beta=10$, and $\mu=-200$ GeV. Contours of tuning are also plotted. The color of the contour indicates the size of $A_t$ for which it applies.}
\end{figure}

In Fig. \ref{fig:fsusyhggratio} we plot the total rate of the $h^0\to\gamma\gamma$ normalized to the SM value as a function of the square averaged stop mass $m^2_T=\frac12(m^2_{\widetilde{T}}+m^2_{\widetilde{t}})$. For definiteness we take the stop soft masses to be equal, $\mu=-200$ GeV, and choose $\tan\beta=10$. The red, blue, and green lines correspond to mixing terms $A_t-\mu\cot\beta$ equal to 100, 400, and 500 GeV respectively. We see that for small mixing the rate is reduced while for larger mixing the rate can be enhanced.

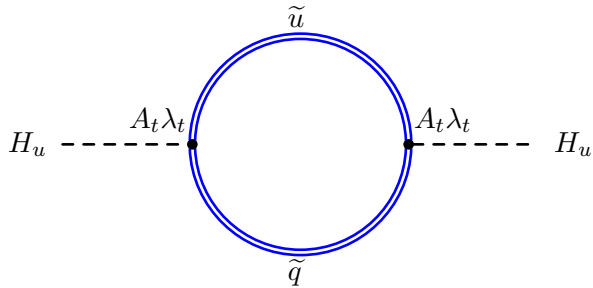
\begin{figure}[t]
  \vspace{5mm}
\begin{tabular}{c}
\begin{fmffile}{fsusyA}
\begin{fmfgraph*}(180,90)
\fmfpen{1.0}
\fmfleft{p1,i1,p2} \fmfright{p3,o1,p4}
\fmfv{l=$H_u$}{i1}\fmfv{l=$H_u$}{o1}
\fmf{dashes}{i1,v1}\fmf{dashes}{v2,o1} 
\fmf{double,right=1,tension=0.3,l.side=left,foreground=blue}{v1,v2,v1}
\fmfv{decor.shape=circle,decor.filled=full,decor.size=1.5thick,l=$A_t\lambda_t$,l.a=115,l.d=5}{v1} \fmfv{decor.shape=circle,decor.filled=full,decor.size=1.5thick,l=$A_t\lambda_t$,l.a=65,l.d=5}{v2}
\fmfv{l=$\widetilde{q}$,l.a=-3,l.d=67}{p1}
\fmfv{l=$\widetilde{u}$,l.a=3,l.d=67}{p2}
\end{fmfgraph*}
\end{fmffile}
\end{tabular}
\caption{\label{fig:stopmixloop} Contribution to the logarithmic divergence in folded SUSY from the stop mixing term.}
\end{figure}

The tuning $\Delta_m$ of the Higgs mass parameter $m^2$ in this model 
differs only slightly from the MSSM case. As in the MTH model, we 
estimate the tuning as
 \begin{equation}
\Delta_m=\left|\frac{2\delta m^2}{m_h^2}\right|^{-1}
 \end{equation} 
 where $\delta m^2$ represents the quantum corrections to the Higgs mass 
parameter and $m_h=$ 125 GeV is the physical Higgs mass. In addition to 
the diagrams in Fig. \ref{fig:fsusyloops}, there is a logarithmic 
divergence due to stop mixing, as shown in Fig. \ref{fig:stopmixloop}. 
From these loops we find, for equal stop soft masses $m_{\text{soft}}$,
 \begin{equation}
|\delta m^2|=\frac{3\lambda_t^2}{16\pi^2}\left[2 m_T^2-2m_t^2-\frac12 m_Z^2\cos 2\beta\left(1+\hat{s}^2 \right) +A_t^2\right]\ln\left(\frac{\Lambda^2}{m_{\text{soft}}^2} \right)
\end{equation}
 where $\Lambda=$ 5 TeV is the cutoff of the model. We have shown the 
tuning for various values of $m_T^2$ in Fig. \ref{fig:fsusyhggratio}. 
The color of each tuning contour corresponds to value of $A_t$ used to 
generate the corresponding curve in the figure.

We see that the modifications to the Higgs couplings in Folded supersymmetry are very small, even when for very mild tuning. Therefore, precision Higgs couplings at the LHC will not strongly constrain naturalness. In this framework, however, top and quark partners are charged under electroweak interaction and will be produced. We therefore briefly investigate the collider limits on F-squarks.

\subsection{Direct searches for F-squarks}

Because the modifications to Higgs rates in folded supersymmetry are small, probes of naturalness in this framework may come from direct searches for F-squarks. Because of the new strong force, collider searches for F-squarks may be complicated by quirky dynamics~\cite{Kang:2008ea}. The quirky narrative for folded SUSY has been outlined in~\cite{Burdman:2008ek}. The most promising signal comes from the production of an up-type and a down-type F-squark through an s-channel $W$. This pair of F-squarks is bound by a quirky string and forms an excited state which loses its excitation energy to soft radiation promptly on collider time scales. The exotic scalar meson, which is now in its ground state, is electrically charged and thus cannot decay into hidden glueballs. In~\cite{Burdman:2008ek} it was shown that the dominant decay of this state is prompt, going to $W\gamma$ with a branching ratio of about 0.85. The predicted signal of this framework is thus a $W\gamma$ resonance at twice the F-squark mass. We will now estimate the current limit  on this framework from an ATLAS $W\gamma$ resonance search~\cite{Aad:2014fha}.

To do this we make some simplifying assumptions. These assumptions lead to a best-case limit, and a more rigorous study is likely to yield weaker bounds. The mass splitting among these two states is expected to be small for the first two generation of F-squarks. Therefore, the time scale $\beta$~decay of one into the other is expected to be longer than the time required for energy loss and decay. We assume that this is the case for the third generation as well.\footnote{If this is not the case, $\beta$~decay will precede the reannihilation of the F-squarks and the dominant channel is a pair of hidden glueballs.} We further assume that the contribution to the $p_T$ of the ground state meson from energy loss is small, which would be the case if the radiation is perfectly isotropic (see~\cite{Burdman:2008ek} for corrections to this approximation). In this case the transverse mass peak is not smeared. Making these assumptions will give us an optimistic estimate for the limit.

\begin{figure}[t]
\centering
\includegraphics[width=0.6\textwidth]{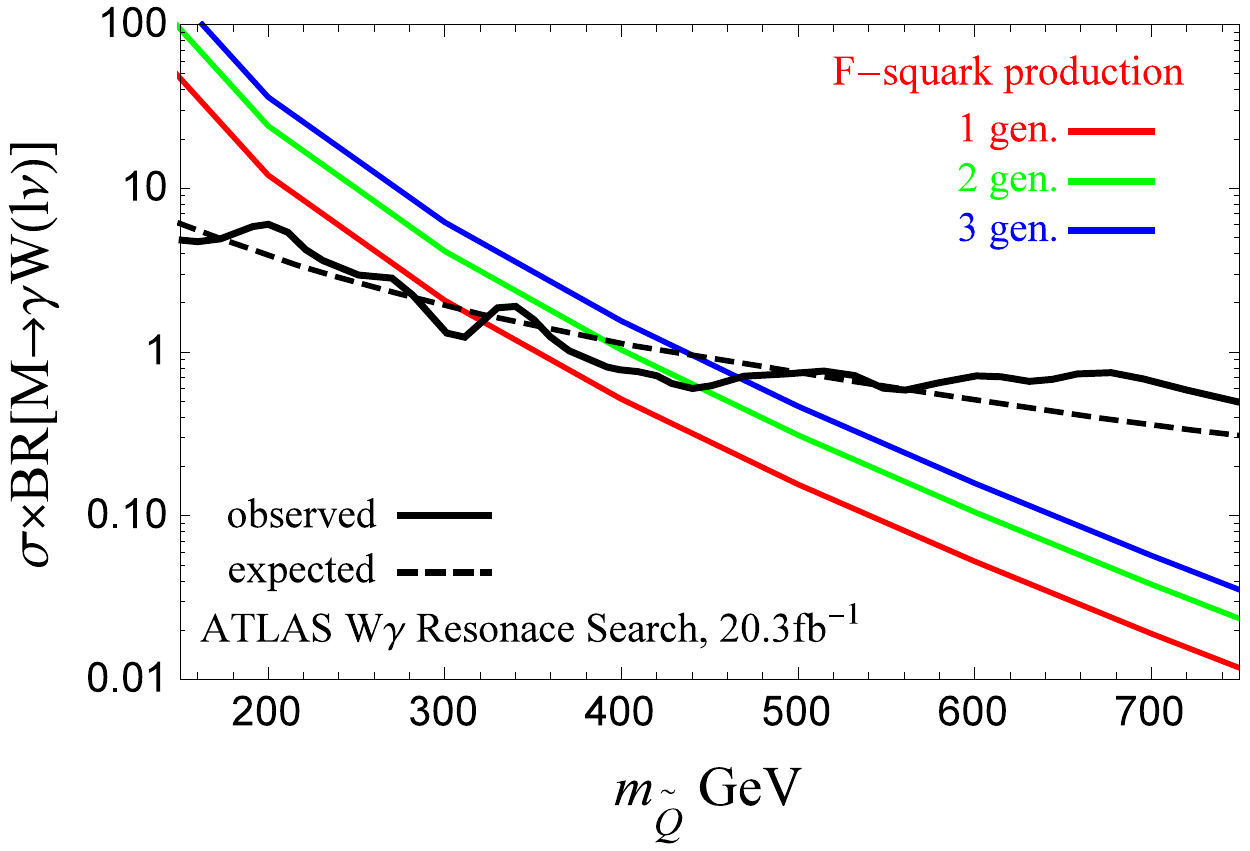}
\caption{\label{fig:squirk-limit} An estimate of the ATLAS limits on the production of an up-down pair of F-squarks  as a function of the F-squark mass, assuming 1, 2, or 3 such generations. }
\end{figure}

The production cross section of the $W\gamma$ resonance is simply the cross section for up-down F-squark pair production. We calculate this cross section  using MadGraph~\cite{Alwall:2011uj} at the 8 TeV LHC. Multiplying by the appropriate branching fractions, we compare this rate to the ATLAS limit in Fig.~\ref{fig:squirk-limit}. We find that the estimated limits on the F-squark mass are about (320, 445, 465) GeV for 1, 2, and 3 generations respectively. 

We conclude that natural models of folded supersymmetry are still allowed by current LHC searches, but future dedicated searches at run-II of the LHC are motivated. We also note that depending on the dominant mechanism of energy loss, the $W\gamma$ resonance may be accompanied by many soft photons contributing to the underlying event~\cite{Harnik:2008ax}.

\section{Quirky Little Higgs\label{sec:qlhiggs}}

\subsection{The Model and Cancellation Mechanism\label{ssec:quirkymodel}}

In Little Higgs models the Higgs doublet emerges as a pNGB whose mass is 
protected against one loop quadratic divergences by collective symmetry 
breaking. To understand how this mechanism operates, consider the 
Simplest Little Higgs model~\cite{Schmaltz:2004de}. In this theory the SU(2$)_{\rm L}\times$U(1$)_{\rm Y}$ gauge symmetry of the SM is embedded in the larger gauge group SU(3$)_W\times$U(1$)_{\rm X}$. All the states in the SM that are doublets under SU(2$)_{\rm L}$ are now promoted to triplets. The 
Higgs sector for this theory is assumed to respect a larger approximate 
global [SU(3)$\times$U(1)]$^2$ symmetry, of which the gauged SU(3$)_W\times$U(1$)_{\rm X}$ is a subgroup. This approximate global symmetry, 
which is explicitly violated by both the gauge and Yukawa interactions, 
is broken to [SU(2)$\times$U(1)]$^2$, which contains SU(2$)_{\rm L}\times$U(1$)_{\rm Y}$ of the SM as a subgroup. The SM Higgs doublet is 
contained among the uneaten pNGBs that emerge from this symmetry 
breaking pattern, and its mass is protected against large radiative 
corrections.

The symmetry breaking pattern may be realized using two scalar triplets 
of SU(3$)_W$, which we denote by $\phi_1$ and $\phi_2$. If the tree level 
potential for these scalars, $V(\phi_1, \phi_2)$ is of the form 
 \begin{equation}
V(\phi_1, \phi_2) = V_1(\phi_1) +  V_2(\phi_2) \; ,
 \end{equation}
 then this sector possesses an [SU(3)$\times$U(1)]$^2$ global 
symmetry. When $\phi_1$ and $\phi_2$ acquire VEVs $f_1$ and $f_2$, this 
symmetry is broken to [SU(2)$\times$U(1)]$^2$. For simplicity we 
assume that the two VEVs are equal, so that $f_1=f_2=f$. However, this 
is not required for the mechanism to work. Of the 10 resulting NGBs, 5 are eaten
while the remaining 5 contain the SM Higgs doublet.

The next step is to understand how the cancellation of quadratic 
divergences associated with the top Yukawa coupling arises in this 
theory. The top sector takes the form
 \begin{equation}
\lambda_1\phi_1 Q t_1+\lambda_2\phi_2 Q t_2\label{eq:linlhtop}
 \end{equation}
 where $Q$ represents the SU(3) triplet containing the third generation 
left-handed quarks, while $t_1$ and $t_2$ are SU(3) singlets that carry 
the same electroweak charge as the right-handed top quark in the SM. 
These interactions do not respect the full [SU(3)$\times$U(1)]$^2$ 
global symmetry but only the gauged SU(3$)_W\times$U(1$)_{\rm X}$ 
subgroup. As a consequence, the potential for $\phi_1$ and $\phi_2$ will 
receive corrections, and the 5 uneaten NGBs will acquire a mass. 
However, as we now explain, this radiatively generated contribution to 
the mass is not quadratically divergent, but only logarithmically 
divergent.

 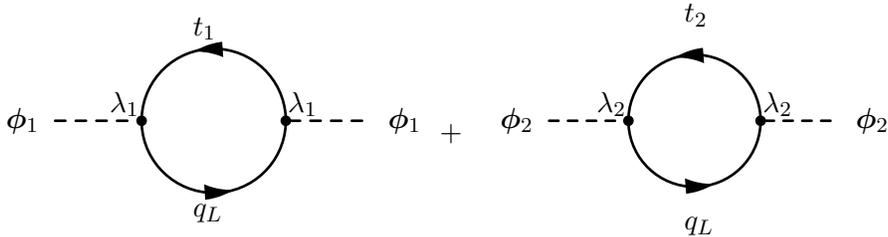
\begin{figure}[ht]
  \vspace{5mm}
\begin{tabular}{ccc}
    \begin{tabular}{c}
\begin{fmffile}{linlhiggs1}
\begin{fmfgraph*}(120,70)
\fmfpen{1.0}
\fmfleft{p1,i1,p2} \fmfright{p3,o1,p4}
\fmfv{l=$\bm{\phi}_1$}{i1}\fmfv{l=$\bm{\phi}_1$}{o1}
\fmf{dashes}{i1,v1}\fmf{dashes}{v2,o1} 
\fmf{fermion,right=1,tension=0.3,l.side=left}{v1,v2,v1}
\fmfv{decor.shape=circle,decor.filled=full,decor.size=1.5thick,l=$\lambda_1$,l.a=115,l.d=2}{v1} \fmfv{decor.shape=circle,decor.filled=full,decor.size=1.5thick,l=$\lambda_1$,l.a=65,l.d=2}{v2}
\fmfv{l=$q_L$,l.a=0,l.d=40}{p1}
\fmfv{l=$t_1$,l.a=0,l.d=40}{p2}
\end{fmfgraph*}
\end{fmffile}
\end{tabular}
\begin{tabular}{c}
\hspace{5mm}$+$ \hspace{5mm}
\end{tabular} 
\begin{tabular}{c}
\begin{fmffile}{linlhiggs2}
\begin{fmfgraph*}(110,80)
\fmfpen{1.0}
\fmfleft{p1,i1,p2} \fmfright{p3,o1,p4}
\fmfv{l=$\bm{\phi}_2$}{i1}\fmfv{l=$\bm{\phi}_2$}{o1}
\fmf{dashes}{i1,v1}\fmf{dashes}{v2,o1} 
\fmf{fermion,right=1,tension=0.3,l.side=left}{v1,v2,v1}
\fmfv{decor.shape=circle,decor.filled=full,decor.size=1.5thick,l=$\lambda_2$,l.a=115,l.d=2}{v1} \fmfv{decor.shape=circle,decor.filled=full,decor.size=1.5thick,l=$\lambda_2$,l.a=65,l.d=2}{v2}
\fmfv{l=$q_L$,l.a=0,l.d=40}{p1}
\fmfv{l=$t_2$,l.a=0,l.d=40}{p2}
\end{fmfgraph*}
\end{fmffile}
\end{tabular}
\end{tabular}
\caption{\label{fig:LHloops} Quadratic divergences from the top sector of the Littlest Higgs model.}
\end{figure}

\begin{figure}[ht]
  \vspace{5mm}
\begin{fmffile}{LHbox}
\begin{fmfgraph*}(65,55)
\fmfpen{1.0}
\fmfleft{i1,i2} \fmfright{o1,o2}\fmfv{l=$\bm{\phi}_1$}{i1}\fmfv{l=$\bm{\phi}_1$}{i2} \fmfv{l=$\bm{\phi}_2$}{o1}\fmfv{l=$\bm{\phi}_2$}{o2}
\fmf{dashes}{i1,v1}\fmf{dashes}{i2,v2} \fmf{dashes}{v3,o1}\fmf{dashes}{v4,o2}
\fmf{fermion,tension=0.2,label=$\bm{q}_L$,l.side=right}{v1,v3} \fmf{fermion,tension=0.2,label=$t_2$,l.side=right}{v3,v4} \fmf{fermion,tension=0.2,label=$\bm{q}_L$,l.side=right}{v4,v2} \fmf{fermion,tension=0.2,label=$t_1$,l.side=right}{v2,v1}
\end{fmfgraph*}
\end{fmffile}
\caption{\label{fig:LHboxloop} Logarithmically divergent contribution to the Higgs potential. This contribution vanishes unless both $\lambda_1$ and $\lambda_2$ are nonzero.}
\end{figure}
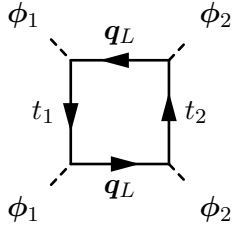

The diagrams that can potentially lead to quadratically divergent 
contributions to the masses of the pNGBs are shown in Fig. \ref{fig:LHloops}. The divergent parts of these graphs are given by
 \begin{equation}
\frac{3}{8\pi^2}\Lambda^2\lambda_1^2\phi_1^{\dag}\phi_1 +\frac{3}{8\pi^2}\Lambda^2\lambda_2^2\phi_2^{\dag}\phi_2\, .
 \end{equation}
 However, we see that these terms respect the full global 
SU(3)$\times$SU(3) symmetry and so cannot contribute to the mass of the 
pNGBs. This is not a coincidence, but a consequence of collective 
symmetry breaking. To see this, note that in \eqref{eq:linlhtop} if 
either of the $\lambda_i$ is set to zero then the Lagrangian for the top 
sector recovers the full SU(3)$\times$SU(3) global symmetry and all the 
resulting NGBs are all massless. We see the global symmetry is violated 
only in the presence of both $\lambda_1$ and $\lambda_2$, which 
collectively break the symmetry. Therefore, any correction to the pNGB 
masses can only arise from a diagram that includes both $\lambda_1$ and 
$\lambda_2$. There are, however, no such quadratically divergent 
diagrams. The lowest order diagram that corrects the potential and 
contains both $\lambda_1$ and $\lambda_2$ is the box diagram, shown in Fig. \ref{fig:LHboxloop}, which is only logarithmically divergent.

\begin{figure}[ht]
  \vspace{5mm}
\begin{tabular}{ccc}
    \begin{tabular}{c}
\begin{fmffile}{lhiggs1}
\begin{fmfgraph*}(120,70)
\fmfpen{1.0}
\fmfleft{p1,i1,p2} \fmfright{p3,o1,p4}
\fmfv{l=$\bm{h}$}{i1}\fmfv{l=$\bm{h}$}{o1}
\fmf{dashes}{i1,v1}\fmf{dashes}{v2,o1} 
\fmf{fermion,right=1,tension=0.3,l.side=left,foreground=blue}{v1,v2,v1}
\fmfv{decor.shape=circle,decor.filled=full,decor.size=1.5thick,l=$\lambda_t,,\lambda_T$,l.a=115,l.d=2}{v1} \fmfv{decor.shape=circle,decor.filled=full,decor.size=1.5thick,l=$\lambda_t,,\lambda_T$,l.a=65,l.d=2}{v2}
\fmfv{l=$q$,l.a=0,l.d=40}{p1}
\fmfv{l=$t^c,,T^c$,l.a=0,l.d=40}{p2}
\end{fmfgraph*}
\end{fmffile}
\end{tabular}
\begin{tabular}{c}
\hspace{5mm}$+$ \hspace{5mm}
\end{tabular} 
\begin{tabular}{c}
\begin{fmffile}{lhiggs2}
\begin{fmfgraph*}(110,80)
\fmfpen{1.0}
\fmfleft{p1,i1,i2,p3} 
\fmfright{p2,o1,o2,p4}
\fmf{phantom}{p3,v2,p4}
\fmf{dashes}{i1,v1,o1}
\fmfv{l=$\bm{h}$}{i1}\fmfv{l=$\bm{h}$}{o1}
\fmffreeze
\fmf{fermion,right,tension=0.1,l.side=right,foreground=blue}{v2,v1}
\fmf{fermion,right,tension=0.1,l.side=right,foreground=blue}{v1,v2}
\fmfv{decor.shape=circle,decor.filled=full,decor.size=1.5thick, l=$-m_T/(2f^2)$,l.a=-90,l.d=5}{v1}
\fmfv{decor.shape=cross,l=$m_T$,l.a=90}{v2}
\fmfv{l=$T^c$,l.d=10,l.a=0}{i2}
\fmfv{l=$T$,l.d=10,l.a=180}{o2}
\end{fmfgraph*}
\end{fmffile}
\end{tabular}
\end{tabular}
\caption{\label{fig:LHquadcancel} Cancellation of quadratic divergences in the Littlest Higgs model. The two fermions must transform under the same SU(3) unless $\lambda_1=\lambda_2$.}
\end{figure}
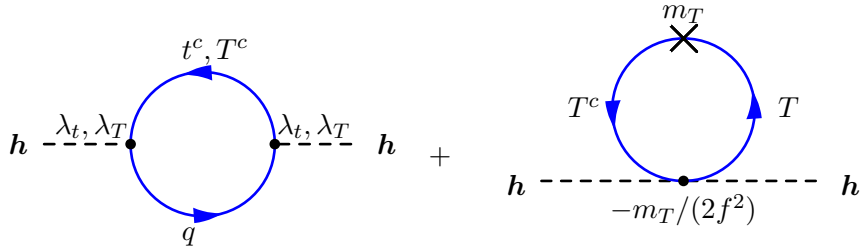

We can show that this protection mechanism depends only on the symmetry 
breaking pattern of the model and is independent of the details of the 
dynamics that breaks the symmetry. To do this, we parametrize the 
uneaten pNGBs, in unitary gauge, by a set of fields $\pi(x)$. It is 
convenient to construct from the $\pi(x)$ two objects $\phi_1$ and 
$\phi_2$ that transform linearly under the full broken SU(3)$\times$SU(3) symmetry.
 \begin{equation}
\begin{array}{cc}
\displaystyle \varphi_1=e^{i\Pi/f}\left(\begin{array}{c}
0\\
0\\
f
\end{array} \right), & \displaystyle \varphi_2=e^{-i\Pi/f}\left(\begin{array}{c}
0\\
0\\
f
\end{array} \right),
\end{array}\label{eq:lhiggsscalars}
\end{equation}
with the relevant degrees of freedom encapsulated by 
\begin{equation}
\Pi=\left(\begin{array}{c|c}
\displaystyle \begin{array}{cc}
0&0\\
0&0
\end{array}& \bm{h}\\
\hline\bm{h}^{\dag}&0
\end{array} \right).
\end{equation} 
 The Lagrangian for the top sector then takes the form
 \begin{equation}
\frac{\lambda_1}{\sqrt{2}}\varphi_1^{\dag}Qt_1+\frac{\lambda_2}{\sqrt{2}} \varphi_2^{\dag} Qt_2\, .
 \end{equation}
 Expanding to quadratic order in $\bm{h}$ and making the definitions
 \begin{align}
t^c&\equiv i\left(\frac{\lambda_1}{\sqrt{\lambda_1^2+\lambda_2^2}}t_2 -\frac{\lambda_1}{\sqrt{\lambda_1^2+\lambda_2^2}}t_1\right)\; ,\\
T^c&\equiv \frac{\lambda_2}{\sqrt{\lambda_1^2+\lambda_2^2}}t_2 +\frac{\lambda_1}{\sqrt{\lambda_1^2+\lambda_2^2}}t_1 
 \end{align}
this becomes
 \begin{equation}
\bm{h}q \left(\lambda_tt^c+\lambda_T T^{c}\right)+m_T TT^c\left(1-\frac{1}{2f^2} \bm{h}^{\dag}\bm{h} \right) \, .
\label{nltop}
\end{equation}
Here we have defined
\begin{equation}
\begin{array}{ccc}
\displaystyle\lambda_t=\frac{\sqrt{2}\lambda_1\lambda_2} {\sqrt{\lambda_1^2+\lambda_2^2}}, &\displaystyle\lambda_T=i\frac{\lambda_2^2-\lambda_1^2} {\sqrt{2}\sqrt{\lambda_1^2+\lambda_2^2}}, &\displaystyle m_T= \frac{f}{\sqrt{2}}\sqrt{\lambda_1^2+\lambda_2^2}\; .
\end{array}
\end{equation} 
 The diagrams contributing to the Higgs mass, see Fig. \ref{fig:LHquadcancel}, demonstrate the cancellation of quadratic divergences. Notice that because $q$ couples to both $t^c$ and $T^c$ that the top partner must transform under the 
same SU(3) as the top. Thus, the two loops have been given the same 
color. If, however, there is some symmetry that forces 
$\lambda_1=\lambda_2$ then the coupling $\lambda_T$ of $q$ to $T^c$ 
vanishes and the cancellation can go through even if $t^c$ and $T^c$ 
transform under different SU(3) color groups.

In Quirky Little Higgs models the one loop quadratic divergences 
generated by the top quark are canceled exactly as in the diagrams shown 
above, but the fermionic top partners $T$ and $T^c$ do not transform 
under the SM color group, SU(3$)_c$. These partners are instead charged 
under a different SU(3), called infracolor, and labeled as SU(3$)_{\rm 
IC}$. However, the electroweak quantum numbers of the quirks are the 
same as those of their SM partners. In this construction, all the 
fermions that are charged under SU(3$)_{\rm IC}$ have masses much above 
the scale where the gauge group gets strong. As a consequence, the 
system exhibits quirky dynamics.

Quirky Little Higgs models can be realized in a 5 dimensional space with 
the extra dimension compactified on $S^1/Z_2$. The breaking of the 
SU(3$)_W\times$U(1$)_X$ gauge group down to the SM is realized by 
boundary conditions and separately by a scalar field $\Phi$ that transforms as 
a triplet under SU(3$)_W$. The 5 dimensional theory also possesses an 
SU(6) gauge symmetry that is broken down to the SM SU(3) color group and 
to SU(3$)_{\rm IC}$ by boundary conditions. This construction allows the 
third generation quark doublet $q$ and the top partner $T$ to emerge as 
zero modes from the same bulk multiplet, but transforming under 
different color groups. The Higgs doublet is contained among the pNGBs 
that emerge from $\Phi$ after the breaking of the SU(3$)_W\times$U(1$)_X$ symmetry. The interactions in \eqref{nltop} arise from 
couplings of $\Phi$ to the multiplets that contain the top quarks and 
the top partners. The SU(6) gauge symmetry ensures the equality of the 
couplings in \eqref{nltop} that is necessary to enforce the cancellation 
of the quadratic divergence.

\subsection{Effects on Higgs Physics\label{eq:hggsquirky}}

When the scalar field $\Phi$ acquires a VEV, the SU(3$)_{W}\times$U(1$)_X$ gauge symmetry is broken down to SU(2$)_{\text{ L}}\times$U(1$)_Y$ of the SM. We associate the SM-like Higgs doublet with some of the NGB modes that emerge from this breaking pattern. We parametrize the relevant degrees of freedom (neglecting the SU(2$)_W$ singlet that plays little role in the phenomenology) as
 \begin{equation}
\Phi=\exp\left(\frac{i}{f}\Pi \right)\left(\begin{array}{c}
0\\
0\\
f
\end{array} \right)
\end{equation}
with
\begin{equation}
\Pi=\left(\begin{array}{cc|c}
0&0&h_1\\
0&0&h_2\\\hline
h_1^{\ast}&h_2^{\ast}&0
\end{array} \right).
\end{equation}
Employing the symbol $\bm{h}$ for the SU(2$)_W$ doublet of $h_1$ and $h_2$ we find
\begin{equation}
\Phi=\left(\begin{array}{c}
\displaystyle\bm{h}\frac{i f}{\sqrt{\bm{h}^{\dag}\bm{h}}} \sin\left(\frac{\sqrt{\bm{h}^{\dag}\bm{h}}}{f} \right)\\
\displaystyle f\cos\left(\frac{\sqrt{\bm{h}^{\dag}\bm{h}}}{f} \right)
\end{array} \right).
\end{equation}

The top sector Yukawa interaction takes the form
\begin{equation}
-i\frac{\lambda_t f}{\sqrt{\bm{h}^{\dag}\bm{h}}}\sin\left(\frac{\sqrt{\bm{h}^{\dag}\bm{h}}}{f} \right)\bm{h}^{\dag}t^cq +\lambda_tf\cos\left(\frac{\sqrt{\bm{h}^{\dag}\bm{h}}}{f} \right)TT^c
\,.
\end{equation}
 After moving to the unitary gauge $h_1=0$, $h_2=(v+\rho)/\sqrt{2}$ this 
becomes
 \begin{equation}
\lambda_t\left[-if\sin\left(\frac{v+\rho}{\sqrt{2}f} \right)t_Lt^c +f\cos\left(\frac{v+\rho}{\sqrt{2}f} \right)TT^c\right]
\end{equation}
 with $t_L$ and $t^c$ transforming under SU(3) color and $T$ and $T^c$ 
transforming under SU(3$)_{\text{IC}}$. Expanding to first order in 
$\rho$ and defining $\vartheta\equiv v/(\sqrt{2}f)$ we find
 \begin{align}
\lambda_t&\left[-i\frac{v_{\text{EW}}}{\sqrt{2}}t_Lt_R\left( 1+\frac{\rho}{v_{\text{EW}}}\cos\vartheta +\ldots \right)\right.\nonumber\\
&\left.\phantom{[]}+f\cos\vartheta TT^c\left(1-\frac{\rho}{v_{EW}}\tan\vartheta \sin\vartheta +\ldots \right) \right]
 \end{align}
 with $v_{\text{EW}}=\sqrt{2}f\sin\vartheta$. We see from this that the 
mass of the top and the mass of the top partner are related by $m_t = 
m_T  \tan \vartheta$. The gauge sector analysis is very similar to 
that of the $A$ sector in MTH models. We expand the gauge kinetic term 
$\left|D_{\mu}\Phi\right|^2$ in the unitary gauge to find the couplings 
between $\rho$ and the gauge bosons:
 \begin{equation}
\left[m^2_{W}W^{+}_{\mu}W^{\mu-}+\frac{m^2_Z}{2}Z_{\mu}Z^{\mu} \right]\left(1+ 2\frac{\rho}{v_{\text{EW}}}\cos\vartheta+\ldots\right).
 \end{equation} 
 We see from this that all zero mode quark and gauge boson couplings are 
suppressed by a universal factor of $\cos\vartheta$ relative to the SM.

\begin{figure}[th]
\centering
\includegraphics[width=0.6\textwidth]{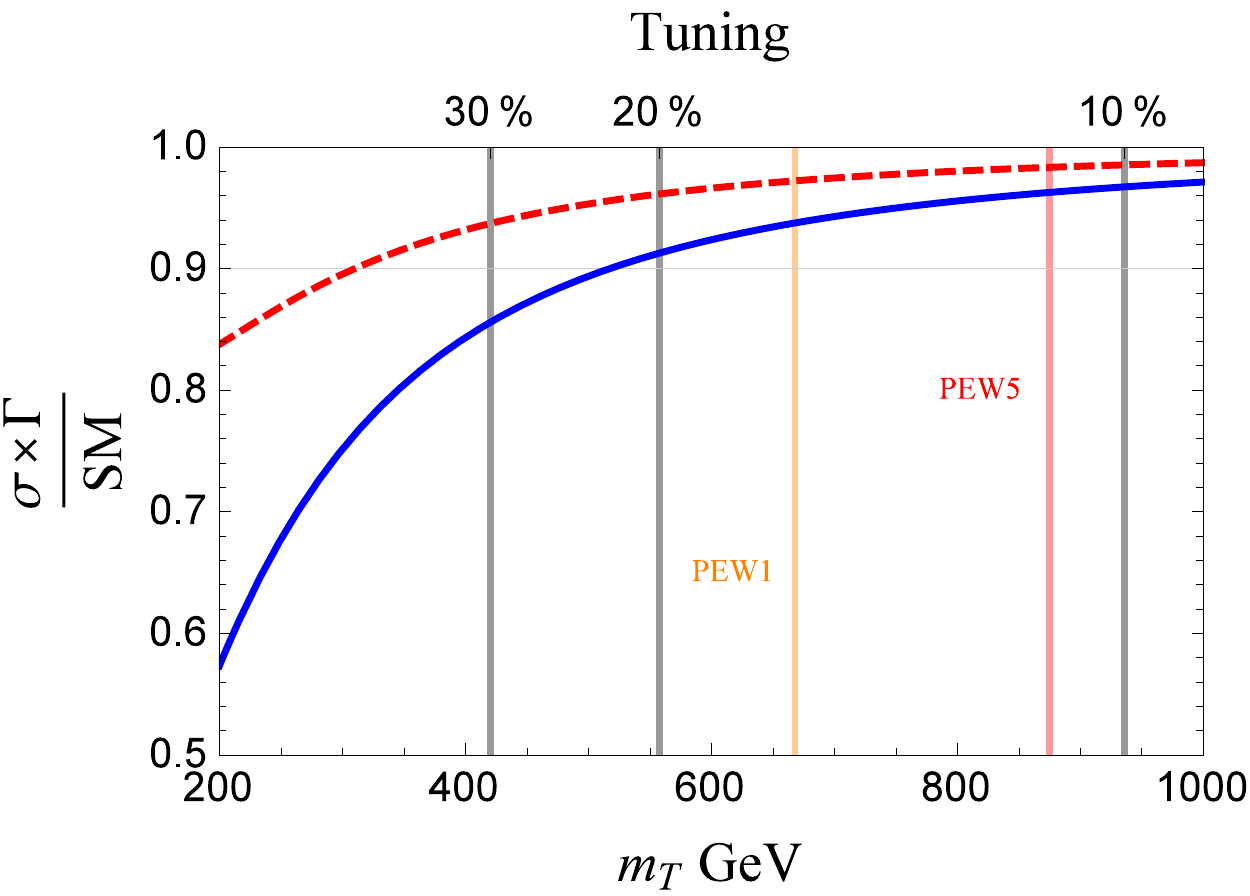}
\caption{\label{fig:QLHrates} Ratios of the rate of Higgs events into a given  final state in Quirky Little Higgs model normalized to the SM. The solid blue line denotes the rates for all final states other than diphoton and the dashed red line denotes the diphoton final state. The vertical orange and red lines represent the 95\% confidence bound from precision electroweak constraints at 1 and 5 TeV respectively. }
\end{figure}

The fact that all the Higgs couplings are corrected by the same factor 
implies that all the production modes are also suppressed by a common 
factor relative to the SM,
 \begin{equation}
\sigma(pp\to \rho)=\cos^2\vartheta \; \sigma_{\text{SM}} (pp\to h).
 \end{equation}
 A similar relation holds for all decay modes of the Higgs 
$\Gamma(\rho\to A_i)$, with the exception of $\Gamma(\rho\to 
\gamma\gamma)$, which receives new contributions from loops involving 
the top partners. The sign of the coupling of the top partner to the 
Higgs is opposite to that of the top. This causes their contributions to 
partially cancel, leading to an enhancement in the $\gamma\gamma$ rate. 
Using Eq. \eqref{eq:genhtogamma} from Appendix A we find
 \begin{align}
\Gamma(\rho\to\gamma\gamma)=\frac{\alpha^2m_{\rho}^3}{1024\pi^3}&\left| \frac{2}{v_{\text{EW}}}\cos \vartheta A_V\left(\frac{4m_W^2}{m_{\rho}^2}\right) +\frac{2}{v_{\text{EW}}} \cos \vartheta\frac43 A_F\left(\frac{4m_t^2}{m_{\rho}^2} \right)\right.\nonumber\\
&\left.\phantom{[]}-\frac{2}{\sqrt{2}f}\tan\vartheta \frac43 A_F\left(\frac{4m_T^2}{m_{\rho}^2}\right)\right|^2.
\end{align}
 We conclude that for all decay modes except the diphoton,
 \begin{equation}
\frac{\sigma(pp\to\rho)\Gamma_{\text{BR}}(\rho\to A_i)} {\sigma_{\text{SM}}(pp\to h)\Gamma_{\text{BR}}^{\text{SM}} (h\to A_i)}= \frac{1}{\displaystyle 1+\frac{m_t^2}{m_T^2}} \; ,
\end{equation}
 where we have neglected tiny effects of order 
$\Gamma(\rho\to\gamma\gamma)/ \Gamma_{\text{SM}}(h)$. For diphoton decays
 \begin{equation} 
\frac{\sigma(pp\to\rho)\Gamma_{\text{BR}}(\rho\to\gamma\gamma)} 
{\sigma_{\text{SM}}(pp\to h)\Gamma_{\text{BR}}^{\text{SM}} (h\to 
\gamma\gamma)}= \frac{\Gamma(\rho \to\gamma\gamma)} 
{\Gamma_{\text{SM}}(h\to\gamma\gamma)} . 
 \end{equation}
 These functions are plotted in Fig. \ref{fig:QLHrates}. The solid blue 
line denotes the rates for all final states other than diphoton and the 
dashed red line denotes the rate to diphotons. Note that even though the 
rate into two photons is enhanced because of the top partner loop, the 
universal suppression factor more than compensates for this, leading to a net suppression.

As with the MTH model, modification of Higgs couplings in the QLH model relative to the SM is constrained by precision electroweak measurements. The 
analysis of the MCHM4 model in \cite{Falkowski:2013dza} also applies to 
the QLH . Their bound on $\epsilon$, where 
$\sqrt{1-\epsilon^2}=\cos\vartheta$, can be translated into a bound on 
the top partner mass. This analysis was carried out assuming a cutoff $\Lambda=$3 TeV. As in the MTH case, we can translate this bound on $\epsilon$ at $\Lambda$ to a bound on $\epsilon '$ at $\Lambda '$; see Eq. \eqref{eq:tuningscaling}. The $2\sigma$ bound on $\epsilon '$ can be translated into a limit on the top partner mass. In Fig. \ref{fig:QLHrates} we denote the bound corresponding to a 1 and 5 TeV cutoff by the vertical orange and red lines respectively. 

Finally, we estimate the tuning $\Delta_m$ of the Higgs mass parameter 
$m^2$ as a function of the top partner mass as a measure of the 
naturalness of the QLH model. We continue to use the formula
 \begin{equation}
\Delta_m=\left|\frac{2\delta m^2}{m_h^2} \right|^{-1}
 \end{equation}
 to estimate the tuning. We have denoted the quantum corrections to the 
Higgs mass parameter as $\delta m^2$ and the physical Higgs mass as 
$m_h=$ 125 GeV.

The diagrams in Fig. \ref{fig:LHquadcancel}, with 
$\lambda_1=\lambda_2=\lambda_t$ lead to
 \begin{equation}
|\delta m^2|=\frac{3\lambda_t^2 m_T^2}{8\pi^2}\ln\left(\frac{\Lambda^2}{m_T^2}\right) \; ,
\end{equation}
 up to finite corrections. We take $\Lambda=$ 5 TeV as the cutoff of the 
theory. In Fig. \ref{fig:QLHrates} we have labeled the top partner 
masses corresponding to 30\%, 20\%, and 10\% tuning. 
We see again that even at the 5\% branching fraction precision expected at full luminosity, the LHC will not be able to probe tunings at the 10\% level. Studies of the direct collider limits on quirky top partners are thus well motivated.

\section{Conclusions\label{sec:con}}

As the LHC bounds on new colored particles continue to grow, theories of 
physics beyond the SM that address the hierarchy problem with colorless 
top partners have become increasingly attractive. Since these new states 
must be light and couple to the Higgs with order one strength to address 
the hierarchy problem, their affects on Higgs production and decay can 
be significant. This suggests the possibility of using precision Higgs 
measurements at the LHC to probe these scenarios.

In this paper we have considered three theories of colorless top 
partners: the Mirror Twin Higgs, Folded Supersymmetry and the Quirky 
Little Higgs. In each case we determined the effects of the top partners 
on Higgs production and decay rates, and used the results to place 
limits on the top partner masses, and therefore on naturalness. We have 
shown that even with 3000 fb$^{-1}$ at 14 TeV, the LHC will not be able to 
strongly disfavor naturalness.

\acknowledgments
We thank Reinard Primulando and Tien-Tien Yu for discussions during early stages of this work. Z.C. and C.V. are supported by the National Science Foundation (NSF) Grant No. PHY-1315155. Fermilab is operated by Fermi Research Alliance, LLC under Contract No. DE-AC02-07CH11359 with the United States Department of Energy. R.H.'s work was supported in part by the NSF under Grant No. PHYS-1066293 and the hospitality of the Aspen Center for Physics. G.B. and L.L. acknowledge the support of the State of S\~{a}o Pulo Research Foundation (FAPESP). G.B. thanks the Brazilian  National Council for Technological and Scientific Development (CNPq) for partial support and the University of Maryland Particle Theory group for its hospitality.

\appendix

\section{General Expressions for the Higgs Decay Rate to Two Photons
\label{sec:genexp}}

In all the models we consider, the effects of new physics on Higgs 
production and decays often occur as simply a multiplicative factor 
relative the SM. In tree level processes this is a reflection of 
modified couplings between the Higgs and SM fields. In loop mediated 
processes, however, we might expect more complicated corrections.

Because we are considering top partners which are not charged under color the gluon fusion and $h\to gg$ decay are affected in exactly the same way as tree level processes. When the top partner is electrically charged, however, the analysis of $h\to \gamma\gamma$ is more subtle. 

At leading order the partial width of the Higgs to $\gamma\gamma$ is given by
\begin{equation}
\Gamma(h\to\gamma\gamma)=\frac{\alpha^2m^3_{h}}{1024\pi^3}\left|\sum\mathcal{M} \right|^2\label{eq:genhtogamma}
\end{equation}
where the amplitudes $\mathcal{M}$ for each electrically charged vector, fermion, or scalar are given by
\begin{align}
\mathcal{M}_V&=\frac{g\left(m_V \right)}{m^2_V}Q^2_VA_V(x_V),\label{eq:vecamp}\\
\mathcal{M}_F&=\frac{g\left(m_F \right)}{m_F^2}Q^2_FA_F(x_F),\label{eq:feramp}\\
\mathcal{M}_S&=\frac{g\left(m_S \right)}{m^2_S}Q^2_SA_S(x_S).\label{eq:scalamp}
\end{align}
In these definitions the $Q$s are the electrical charges in units of $e$, the charge of the proton and $g(m)$ is the couping of the particle to the Higgs. The $A$ functions are given by
\begin{align}
A_V(x)&=-x^2\left[\frac{2}{x^2}+\frac{3}{x}+3\left(\frac{2}{x}-1 \right)\arcsin^2\left(\frac{1}{\sqrt{x}} \right) \right],\label{eq:vecfunc}\\
A_F(x)&=2x^2\left[\frac{1}{x}+\left(\frac{1}{x}-1 \right)\arcsin^2\left(\frac{1}{\sqrt{x}} \right) \right],\label{eq:ferfunc}\\
A_S(x)&=-x^2\left[\frac{1}{x}-\arcsin^2\left(\frac{1}{\sqrt{x}} \right) \right]\label{eq:scalfunc}
\end{align}
where $x_i=4m_i^2/m_h^2$ and is understood to be greater than one. The couplings $g$ are defined by
\begin{equation}
\frac{g(m)}{m^2}=\frac{1}{m^2(v)}\frac{\partial m^2(v)}{\partial v}
\end{equation}
where in the case of fermions the mass squared is taken to mean $|m(v)|^2$.

\section{MSSM with Extra U(1$)_X$ \label{sec:mssmu1}}
In this appendix we add a U(1$)_X$ gauge symmetry, with coupling $g_X$, to the MSSM which is then spontaneously broken. This affects the Higgs mass, the stop masses, and the Higgs couplings to the stops. We follow closely the work of~\cite{Batra:2003nj}.

All MSSM matter content is given equal charge under hypercharge and U(1$)_X$. In addition, the heavy scalar fields $\phi$ and $\phi^c$, which spontaneously break the symmetry, carry charges $\pm q$ under the new U(1$)_X$ but are singlets under every other MSSM gauge group. These fields are part of chiral superfields $\Phi$ and $\Phi^c$ with superpotential
\begin{equation}
\mathcal{W}=\lambda S\left(\Phi\Phi^c-w^2 \right)\label{eq:superpot}
\end{equation}
and soft masses
\begin{equation}
m_{\phi}^2\left(|\phi|^2+|\phi^c|^2 \right).
\end{equation}
For $\lambda^2w^2>m_{\phi}^2$ and equal soft masses these scalars obtain identical nonzero VEVs $\langle\phi\rangle$. The U(1$)_X$ gauge field $Z_{\mu}'$ also gets a mass $m_{Z'}=2qg_X\langle\phi\rangle$. 

The usual MSSM $D$-terms
\begin{equation}
\frac{g_L^2}{2}\left(\sum_{\text{MSSM}}\phi_i^{\ast}q_i\sigma^a\phi_i \right)^2 +\frac{g^2_Y}{2}\left(\sum_{\text{MSSM}}\phi_i^{\ast}q_i\phi_i \right)^2 
\end{equation}
(with the $q_i$ denoting the charge of the $i$th field with respect to the appropriate gauge symmetry) are joined by
\begin{equation}
\frac{g^2_X}{2}\left(\sum_{\text{MSSM}}\phi_i^{\ast}q_i\phi_i +q|\phi|^2-q|\phi^c|^2 \right)^2.
\end{equation}
When $\phi$ and $\phi^c$ have masses much higher than the weak scale we can integrate them out. This generates the leading $D$-terms
\begin{equation}
\frac{g_L^2}{2}\left(\sum_{\text{MSSM}}\phi_i^{\ast}q_i\sigma^a\phi_i \right)^2 +\frac{g^2_Y+\hat{g}^2}{2}\left(\sum_{\text{MSSM}}\phi_i^{\ast}q_i\phi_i \right)^2 \label{eq:newdterm}
\end{equation}
where
\begin{equation}
\hat{s}^2=g_X^2\left(1+\frac{m_{Z'}^2}{2m_{\phi}^2} \right)^{-1}\frac{ v_{\text{EW}}^2}{4m_{Z}^2}.
\end{equation}

This effective enhancement of the hypercharge $D$-term raises the tree level Higgs mass to
\begin{equation}
m_{h^0}^2=m_Z^2\cos^22\beta\left(1+\hat{s}^2 \right).
\end{equation}
The $D$-term contributions to the Higgs-stop couplings and the stop masses are similarly modified, as shown in the body of the paper. All numerical results, see Fig. \ref{fig:fsusyhggratio}, use the value of $\hat{s}$ such that $m_{h^0}=125$ GeV with stop loop corrections to the Higgs mass included\cite{Martin:1997ns}:
\begin{align}
m_{h^0}^2=m_Z^2\cos^22\beta\left(1+\hat{s}^2 \right)+&\frac{3\lambda_t^2\sin^2\beta}{2\pi^2}\left\{m_t^2\ln\left( \frac{m_{\widetilde{T}}m_{\widetilde{t}}}{m_t^2}\right)+\frac{\sin^22\alpha_t} {4}(m_{\widetilde{T}}^2-m_{\widetilde{t}}^2)\ln\left(\frac{m_{\widetilde{T}}^2} {m_{\widetilde{t}}^2} \right)\right.\nonumber\\
&\left.+\frac{\sin^42\alpha_t}{16m_t^2}\left[(m_{\widetilde{T}}^2 -m_{\widetilde{t}}^2)^2-\frac12(m_{\widetilde{T}}^4-m_{\widetilde{t}}^4)\ln\left( \frac{m_{\widetilde{T}}^2} {m_{\widetilde{t}}^2}\right) \right]\right\}
\end{align}
where we have used the definition of $\sin 2\alpha_t$ from \eqref{eq:stopangle}.

\bibliography{colorlesstopbib}

\begin{thebibliography}{39}%
\makeatletter
\providecommand \@ifxundefined [1]{%
 \@ifx{#1\undefined}
}%
\providecommand \@ifnum [1]{%
 \ifnum #1\expandafter \@firstoftwo
 \else \expandafter \@secondoftwo
 \fi
}%
\providecommand \@ifx [1]{%
 \ifx #1\expandafter \@firstoftwo
 \else \expandafter \@secondoftwo
 \fi
}%
\providecommand \natexlab [1]{#1}%
\providecommand \enquote  [1]{``#1''}%
\providecommand \bibnamefont  [1]{#1}%
\providecommand \bibfnamefont [1]{#1}%
\providecommand \citenamefont [1]{#1}%
\providecommand \href@noop [0]{\@secondoftwo}%
\providecommand \href [0]{\begingroup \@sanitize@url \@href}%
\providecommand \@href[1]{\@@startlink{#1}\@@href}%
\providecommand \@@href[1]{\endgroup#1\@@endlink}%
\providecommand \@sanitize@url [0]{\catcode `\\12\catcode `\$12\catcode
  `\&12\catcode `\#12\catcode `\^12\catcode `\_12\catcode `\%12\relax}%
\providecommand \@@startlink[1]{}%
\providecommand \@@endlink[0]{}%
\providecommand \url  [0]{\begingroup\@sanitize@url \@url }%
\providecommand \@url [1]{\endgroup\@href {#1}{\urlprefix }}%
\providecommand \urlprefix  [0]{URL }%
\providecommand \Eprint [0]{\href }%
\providecommand \doibase [0]{http://dx.doi.org/}%
\providecommand \selectlanguage [0]{\@gobble}%
\providecommand \bibinfo  [0]{\@secondoftwo}%
\providecommand \bibfield  [0]{\@secondoftwo}%
\providecommand \translation [1]{[#1]}%
\providecommand \BibitemOpen [0]{}%
\providecommand \bibitemStop [0]{}%
\providecommand \bibitemNoStop [0]{.\EOS\space}%
\providecommand \EOS [0]{\spacefactor3000\relax}%
\providecommand \BibitemShut  [1]{\csname bibitem#1\endcsname}%
\let\auto@bib@innerbib\@empty
\bibitem [{\citenamefont {Aad}\ \emph {et~al.}(2012)\citenamefont {Aad} \emph
  {et~al.}}]{Aad:2012tfa}%
  \BibitemOpen
  \bibfield  {author} {\bibinfo {author} {\bibfnamefont {G.}~\bibnamefont
  {Aad}} \emph {et~al.} (\bibinfo {collaboration} {ATLAS Collaboration}),\
  }\href {\doibase 10.1016/j.physletb.2012.08.020} {\bibfield  {journal}
  {\bibinfo  {journal} {Phys.Lett.}\ }\textbf {\bibinfo {volume} {B716}},\
  \bibinfo {pages} {1} (\bibinfo {year} {2012})},\ \Eprint
  {http://arxiv.org/abs/1207.7214} {arXiv:1207.7214 [hep-ex]} \BibitemShut
  {NoStop}%
\bibitem [{\citenamefont {Chatrchyan}\ \emph {et~al.}(2012)\citenamefont
  {Chatrchyan} \emph {et~al.}}]{Chatrchyan:2012ufa}%
  \BibitemOpen
  \bibfield  {author} {\bibinfo {author} {\bibfnamefont {S.}~\bibnamefont
  {Chatrchyan}} \emph {et~al.} (\bibinfo {collaboration} {CMS Collaboration}),\
  }\href {\doibase 10.1016/j.physletb.2012.08.021} {\bibfield  {journal}
  {\bibinfo  {journal} {Phys.Lett.}\ }\textbf {\bibinfo {volume} {B716}},\
  \bibinfo {pages} {30} (\bibinfo {year} {2012})},\ \Eprint
  {http://arxiv.org/abs/1207.7235} {arXiv:1207.7235 [hep-ex]} \BibitemShut
  {NoStop}%
\bibitem [{\citenamefont {Martin}(1997)}]{Martin:1997ns}%
  \BibitemOpen
  \bibfield  {author} {\bibinfo {author} {\bibfnamefont {S.~P.}\ \bibnamefont
  {Martin}},\ }\href@noop {} {\  (\bibinfo {year} {1997})},\ \Eprint
  {http://arxiv.org/abs/hep-ph/9709356} {arXiv:hep-ph/9709356 [hep-ph]}
  \BibitemShut {NoStop}%
\bibitem [{\citenamefont {Arkani-Hamed}\ \emph {et~al.}(2001)\citenamefont
  {Arkani-Hamed}, \citenamefont {Cohen},\ and\ \citenamefont
  {Georgi}}]{ArkaniHamed:2001nc}%
  \BibitemOpen
  \bibfield  {author} {\bibinfo {author} {\bibfnamefont {N.}~\bibnamefont
  {Arkani-Hamed}}, \bibinfo {author} {\bibfnamefont {A.~G.}\ \bibnamefont
  {Cohen}}, \ and\ \bibinfo {author} {\bibfnamefont {H.}~\bibnamefont
  {Georgi}},\ }\href {\doibase 10.1016/S0370-2693(01)00741-9} {\bibfield
  {journal} {\bibinfo  {journal} {Phys.Lett.}\ }\textbf {\bibinfo {volume}
  {B513}},\ \bibinfo {pages} {232} (\bibinfo {year} {2001})},\ \Eprint
  {http://arxiv.org/abs/hep-ph/0105239} {arXiv:hep-ph/0105239 [hep-ph]}
  \BibitemShut {NoStop}%
\bibitem [{\citenamefont {Arkani-Hamed}\ \emph
  {et~al.}(2002{\natexlab{a}})\citenamefont {Arkani-Hamed}, \citenamefont
  {Cohen}, \citenamefont {Katz}, \citenamefont {Nelson}, \citenamefont
  {Gregoire} \emph {et~al.}}]{ArkaniHamed:2002qx}%
  \BibitemOpen
  \bibfield  {author} {\bibinfo {author} {\bibfnamefont {N.}~\bibnamefont
  {Arkani-Hamed}}, \bibinfo {author} {\bibfnamefont {A.}~\bibnamefont {Cohen}},
  \bibinfo {author} {\bibfnamefont {E.}~\bibnamefont {Katz}}, \bibinfo {author}
  {\bibfnamefont {A.}~\bibnamefont {Nelson}}, \bibinfo {author} {\bibfnamefont
  {T.}~\bibnamefont {Gregoire}},  \emph {et~al.},\ }\href {\doibase
  10.1088/1126-6708/2002/08/021} {\bibfield  {journal} {\bibinfo  {journal}
  {JHEP}\ }\textbf {\bibinfo {volume} {0208}},\ \bibinfo {pages} {021}
  (\bibinfo {year} {2002}{\natexlab{a}})},\ \Eprint
  {http://arxiv.org/abs/hep-ph/0206020} {arXiv:hep-ph/0206020 [hep-ph]}
  \BibitemShut {NoStop}%
\bibitem [{\citenamefont {Arkani-Hamed}\ \emph
  {et~al.}(2002{\natexlab{b}})\citenamefont {Arkani-Hamed}, \citenamefont
  {Cohen}, \citenamefont {Katz},\ and\ \citenamefont
  {Nelson}}]{ArkaniHamed:2002qy}%
  \BibitemOpen
  \bibfield  {author} {\bibinfo {author} {\bibfnamefont {N.}~\bibnamefont
  {Arkani-Hamed}}, \bibinfo {author} {\bibfnamefont {A.}~\bibnamefont {Cohen}},
  \bibinfo {author} {\bibfnamefont {E.}~\bibnamefont {Katz}}, \ and\ \bibinfo
  {author} {\bibfnamefont {A.}~\bibnamefont {Nelson}},\ }\href {\doibase
  10.1088/1126-6708/2002/07/034} {\bibfield  {journal} {\bibinfo  {journal}
  {JHEP}\ }\textbf {\bibinfo {volume} {0207}},\ \bibinfo {pages} {034}
  (\bibinfo {year} {2002}{\natexlab{b}})},\ \Eprint
  {http://arxiv.org/abs/hep-ph/0206021} {arXiv:hep-ph/0206021 [hep-ph]}
  \BibitemShut {NoStop}%
\bibitem [{\citenamefont {Schmaltz}(2004)}]{Schmaltz:2004de}%
  \BibitemOpen
  \bibfield  {author} {\bibinfo {author} {\bibfnamefont {M.}~\bibnamefont
  {Schmaltz}},\ }\href {\doibase 10.1088/1126-6708/2004/08/056} {\bibfield
  {journal} {\bibinfo  {journal} {JHEP}\ }\textbf {\bibinfo {volume} {0408}},\
  \bibinfo {pages} {056} (\bibinfo {year} {2004})},\ \Eprint
  {http://arxiv.org/abs/hep-ph/0407143} {arXiv:hep-ph/0407143 [hep-ph]}
  \BibitemShut {NoStop}%
\bibitem [{\citenamefont {Schmaltz}\ and\ \citenamefont
  {Tucker-Smith}(2005)}]{Schmaltz:2005ky}%
  \BibitemOpen
  \bibfield  {author} {\bibinfo {author} {\bibfnamefont {M.}~\bibnamefont
  {Schmaltz}}\ and\ \bibinfo {author} {\bibfnamefont {D.}~\bibnamefont
  {Tucker-Smith}},\ }\href {\doibase 10.1146/annurev.nucl.55.090704.151502}
  {\bibfield  {journal} {\bibinfo  {journal} {Ann.Rev.Nucl.Part.Sci.}\ }\textbf
  {\bibinfo {volume} {55}},\ \bibinfo {pages} {229} (\bibinfo {year} {2005})},\
  \Eprint {http://arxiv.org/abs/hep-ph/0502182} {arXiv:hep-ph/0502182 [hep-ph]}
  \BibitemShut {NoStop}%
\bibitem [{\citenamefont {Chatrchyan}\ \emph {et~al.}(2013)\citenamefont
  {Chatrchyan} \emph {et~al.}}]{Chatrchyan:2013xna}%
  \BibitemOpen
  \bibfield  {author} {\bibinfo {author} {\bibfnamefont {S.}~\bibnamefont
  {Chatrchyan}} \emph {et~al.} (\bibinfo {collaboration} {CMS Collaboration}),\
  }\href {\doibase 10.1140/epjc/s10052-013-2677-2} {\bibfield  {journal}
  {\bibinfo  {journal} {Eur.Phys.J.}\ }\textbf {\bibinfo {volume} {C73}},\
  \bibinfo {pages} {2677} (\bibinfo {year} {2013})},\ \Eprint
  {http://arxiv.org/abs/1308.1586} {arXiv:1308.1586 [hep-ex]} \BibitemShut
  {NoStop}%
\bibitem [{\citenamefont {{CMS-PAS-SUS-14-011}}(2014)}]{CMS:2014wsa}%
  \BibitemOpen
  \bibfield  {author} {\bibinfo {author} {\bibnamefont {{CMS-PAS-SUS-14-011}}}
  (\bibinfo {collaboration} {CMS Collaboration}),\ }\href@noop {} {\  (\bibinfo
  {year} {2014})}\BibitemShut {NoStop}%
\bibitem [{\citenamefont {Aad}\ \emph {et~al.}(2014{\natexlab{a}})\citenamefont
  {Aad} \emph {et~al.}}]{Aad:2014kra}%
  \BibitemOpen
  \bibfield  {author} {\bibinfo {author} {\bibfnamefont {G.}~\bibnamefont
  {Aad}} \emph {et~al.} (\bibinfo {collaboration} {ATLAS Collaboration}),\
  }\href@noop {} {\  (\bibinfo {year} {2014}{\natexlab{a}})},\ \Eprint
  {http://arxiv.org/abs/1407.0583} {arXiv:1407.0583 [hep-ex]} \BibitemShut
  {NoStop}%
\bibitem [{\citenamefont {Aad}\ \emph {et~al.}(2014{\natexlab{b}})\citenamefont
  {Aad} \emph {et~al.}}]{Aad:2014bva}%
  \BibitemOpen
  \bibfield  {author} {\bibinfo {author} {\bibfnamefont {G.}~\bibnamefont
  {Aad}} \emph {et~al.} (\bibinfo {collaboration} {ATLAS Collaboration}),\
  }\href {\doibase 10.1007/JHEP09(2014)015} {\bibfield  {journal} {\bibinfo
  {journal} {JHEP}\ }\textbf {\bibinfo {volume} {1409}},\ \bibinfo {pages}
  {015} (\bibinfo {year} {2014}{\natexlab{b}})},\ \Eprint
  {http://arxiv.org/abs/1406.1122} {arXiv:1406.1122 [hep-ex]} \BibitemShut
  {NoStop}%
\bibitem [{\citenamefont {Martin}(2007)}]{Martin:2007gf}%
  \BibitemOpen
  \bibfield  {author} {\bibinfo {author} {\bibfnamefont {S.~P.}\ \bibnamefont
  {Martin}},\ }\href {\doibase 10.1103/PhysRevD.75.115005} {\bibfield
  {journal} {\bibinfo  {journal} {Phys.Rev.}\ }\textbf {\bibinfo {volume}
  {D75}},\ \bibinfo {pages} {115005} (\bibinfo {year} {2007})},\ \Eprint
  {http://arxiv.org/abs/hep-ph/0703097} {arXiv:hep-ph/0703097 [HEP-PH]}
  \BibitemShut {NoStop}%
\bibitem [{\citenamefont {Fan}\ \emph {et~al.}(2011)\citenamefont {Fan},
  \citenamefont {Reece},\ and\ \citenamefont {Ruderman}}]{Fan:2011yu}%
  \BibitemOpen
  \bibfield  {author} {\bibinfo {author} {\bibfnamefont {J.}~\bibnamefont
  {Fan}}, \bibinfo {author} {\bibfnamefont {M.}~\bibnamefont {Reece}}, \ and\
  \bibinfo {author} {\bibfnamefont {J.~T.}\ \bibnamefont {Ruderman}},\ }\href
  {\doibase 10.1007/JHEP11(2011)012} {\bibfield  {journal} {\bibinfo  {journal}
  {JHEP}\ }\textbf {\bibinfo {volume} {1111}},\ \bibinfo {pages} {012}
  (\bibinfo {year} {2011})},\ \Eprint {http://arxiv.org/abs/1105.5135}
  {arXiv:1105.5135 [hep-ph]} \BibitemShut {NoStop}%
\bibitem [{\citenamefont {Chacko}\ \emph
  {et~al.}(2006{\natexlab{a}})\citenamefont {Chacko}, \citenamefont {Goh},\
  and\ \citenamefont {Harnik}}]{Twnhggs2006}%
  \BibitemOpen
  \bibfield  {author} {\bibinfo {author} {\bibfnamefont {Z.}~\bibnamefont
  {Chacko}}, \bibinfo {author} {\bibfnamefont {H.-S.}\ \bibnamefont {Goh}}, \
  and\ \bibinfo {author} {\bibfnamefont {R.}~\bibnamefont {Harnik}},\ }\href
  {\doibase 10.1103/PhysRevLett.96.231802} {\bibfield  {journal} {\bibinfo
  {journal} {Phys.Rev.Lett.}\ }\textbf {\bibinfo {volume} {96}},\ \bibinfo
  {pages} {231802} (\bibinfo {year} {2006}{\natexlab{a}})},\ \Eprint
  {http://arxiv.org/abs/hep-ph/0506256} {arXiv:hep-ph/0506256 [hep-ph]}
  \BibitemShut {NoStop}%
\bibitem [{\citenamefont {Burdman}\ \emph {et~al.}(2007)\citenamefont
  {Burdman}, \citenamefont {Chacko}, \citenamefont {Goh},\ and\ \citenamefont
  {Harnik}}]{fldsusy2007}%
  \BibitemOpen
  \bibfield  {author} {\bibinfo {author} {\bibfnamefont {G.}~\bibnamefont
  {Burdman}}, \bibinfo {author} {\bibfnamefont {Z.}~\bibnamefont {Chacko}},
  \bibinfo {author} {\bibfnamefont {H.-S.}\ \bibnamefont {Goh}}, \ and\
  \bibinfo {author} {\bibfnamefont {R.}~\bibnamefont {Harnik}},\ }\href
  {\doibase 10.1088/1126-6708/2007/02/009} {\bibfield  {journal} {\bibinfo
  {journal} {JHEP}\ }\textbf {\bibinfo {volume} {02}},\ \bibinfo {pages} {009}
  (\bibinfo {year} {2007})},\ \Eprint {http://arxiv.org/abs/hep-ph/0609152}
  {arXiv:hep-ph/0609152 [hep-ph]} \BibitemShut {NoStop}%
\bibitem [{\citenamefont {Poland}\ and\ \citenamefont
  {Thaler}(2008)}]{Poland:2008ev}%
  \BibitemOpen
  \bibfield  {author} {\bibinfo {author} {\bibfnamefont {D.}~\bibnamefont
  {Poland}}\ and\ \bibinfo {author} {\bibfnamefont {J.}~\bibnamefont
  {Thaler}},\ }\href {\doibase 10.1088/1126-6708/2008/11/083} {\bibfield
  {journal} {\bibinfo  {journal} {JHEP}\ }\textbf {\bibinfo {volume} {0811}},\
  \bibinfo {pages} {083} (\bibinfo {year} {2008})},\ \Eprint
  {http://arxiv.org/abs/0808.1290} {arXiv:0808.1290 [hep-ph]} \BibitemShut
  {NoStop}%
\bibitem [{\citenamefont {Cai}\ \emph {et~al.}(2009)\citenamefont {Cai},
  \citenamefont {Cheng},\ and\ \citenamefont {Terning}}]{Quirkyhggs2009}%
  \BibitemOpen
  \bibfield  {author} {\bibinfo {author} {\bibfnamefont {H.}~\bibnamefont
  {Cai}}, \bibinfo {author} {\bibfnamefont {H.-C.}\ \bibnamefont {Cheng}}, \
  and\ \bibinfo {author} {\bibfnamefont {J.}~\bibnamefont {Terning}},\ }\href
  {\doibase 10.1088/1126-6708/2009/05/045} {\bibfield  {journal} {\bibinfo
  {journal} {JHEP}\ }\textbf {\bibinfo {volume} {05}},\ \bibinfo {pages} {045}
  (\bibinfo {year} {2009})},\ \Eprint {http://arxiv.org/abs/0812.0843}
  {arXiv:0812.0843 [hep-ph]} \BibitemShut {NoStop}%
\bibitem [{\citenamefont {Barbieri}\ \emph {et~al.}(2005)\citenamefont
  {Barbieri}, \citenamefont {Gregoire},\ and\ \citenamefont
  {Hall}}]{Barbieri:2005ri}%
  \BibitemOpen
  \bibfield  {author} {\bibinfo {author} {\bibfnamefont {R.}~\bibnamefont
  {Barbieri}}, \bibinfo {author} {\bibfnamefont {T.}~\bibnamefont {Gregoire}},
  \ and\ \bibinfo {author} {\bibfnamefont {L.~J.}\ \bibnamefont {Hall}},\
  }\href@noop {} {\  (\bibinfo {year} {2005})},\ \Eprint
  {http://arxiv.org/abs/hep-ph/0509242} {arXiv:hep-ph/0509242 [hep-ph]}
  \BibitemShut {NoStop}%
\bibitem [{\citenamefont {Chacko}\ \emph
  {et~al.}(2006{\natexlab{b}})\citenamefont {Chacko}, \citenamefont {Nomura},
  \citenamefont {Papucci},\ and\ \citenamefont {Perez}}]{Chacko:2005vw}%
  \BibitemOpen
  \bibfield  {author} {\bibinfo {author} {\bibfnamefont {Z.}~\bibnamefont
  {Chacko}}, \bibinfo {author} {\bibfnamefont {Y.}~\bibnamefont {Nomura}},
  \bibinfo {author} {\bibfnamefont {M.}~\bibnamefont {Papucci}}, \ and\
  \bibinfo {author} {\bibfnamefont {G.}~\bibnamefont {Perez}},\ }\href
  {\doibase 10.1088/1126-6708/2006/01/126} {\bibfield  {journal} {\bibinfo
  {journal} {JHEP}\ }\textbf {\bibinfo {volume} {0601}},\ \bibinfo {pages}
  {126} (\bibinfo {year} {2006}{\natexlab{b}})},\ \Eprint
  {http://arxiv.org/abs/hep-ph/0510273} {arXiv:hep-ph/0510273 [hep-ph]}
  \BibitemShut {NoStop}%
\bibitem [{\citenamefont {Chang}\ \emph {et~al.}(2007)\citenamefont {Chang},
  \citenamefont {Hall},\ and\ \citenamefont {Weiner}}]{Chang:2006ra}%
  \BibitemOpen
  \bibfield  {author} {\bibinfo {author} {\bibfnamefont {S.}~\bibnamefont
  {Chang}}, \bibinfo {author} {\bibfnamefont {L.~J.}\ \bibnamefont {Hall}}, \
  and\ \bibinfo {author} {\bibfnamefont {N.}~\bibnamefont {Weiner}},\ }\href
  {\doibase 10.1103/PhysRevD.75.035009} {\bibfield  {journal} {\bibinfo
  {journal} {Phys.Rev.}\ }\textbf {\bibinfo {volume} {D75}},\ \bibinfo {pages}
  {035009} (\bibinfo {year} {2007})},\ \Eprint
  {http://arxiv.org/abs/hep-ph/0604076} {arXiv:hep-ph/0604076 [hep-ph]}
  \BibitemShut {NoStop}%
\bibitem [{\citenamefont {Craig}\ and\ \citenamefont
  {Howe}(2014)}]{Craig:2013fga}%
  \BibitemOpen
  \bibfield  {author} {\bibinfo {author} {\bibfnamefont {N.}~\bibnamefont
  {Craig}}\ and\ \bibinfo {author} {\bibfnamefont {K.}~\bibnamefont {Howe}},\
  }\href {\doibase 10.1007/JHEP03(2014)140} {\bibfield  {journal} {\bibinfo
  {journal} {JHEP}\ }\textbf {\bibinfo {volume} {1403}},\ \bibinfo {pages}
  {140} (\bibinfo {year} {2014})},\ \Eprint {http://arxiv.org/abs/1312.1341}
  {arXiv:1312.1341 [hep-ph]} \BibitemShut {NoStop}%
\bibitem [{\citenamefont {Craig}\ \emph {et~al.}(2014)\citenamefont {Craig},
  \citenamefont {Knapen},\ and\ \citenamefont {Longhi}}]{Craig:2014aea}%
  \BibitemOpen
  \bibfield  {author} {\bibinfo {author} {\bibfnamefont {N.}~\bibnamefont
  {Craig}}, \bibinfo {author} {\bibfnamefont {S.}~\bibnamefont {Knapen}}, \
  and\ \bibinfo {author} {\bibfnamefont {P.}~\bibnamefont {Longhi}},\
  }\href@noop {} {\  (\bibinfo {year} {2014})},\ \Eprint
  {http://arxiv.org/abs/1410.6808} {arXiv:1410.6808 [hep-ph]} \BibitemShut
  {NoStop}%
\bibitem [{\citenamefont {Strassler}\ and\ \citenamefont
  {Zurek}(2007)}]{Strassler:2006im}%
  \BibitemOpen
  \bibfield  {author} {\bibinfo {author} {\bibfnamefont {M.~J.}\ \bibnamefont
  {Strassler}}\ and\ \bibinfo {author} {\bibfnamefont {K.~M.}\ \bibnamefont
  {Zurek}},\ }\href {\doibase 10.1016/j.physletb.2007.06.055} {\bibfield
  {journal} {\bibinfo  {journal} {Phys.Lett.}\ }\textbf {\bibinfo {volume}
  {B651}},\ \bibinfo {pages} {374} (\bibinfo {year} {2007})},\ \Eprint
  {http://arxiv.org/abs/hep-ph/0604261} {arXiv:hep-ph/0604261 [hep-ph]}
  \BibitemShut {NoStop}%
\bibitem [{\citenamefont {Kang}\ and\ \citenamefont
  {Luty}(2009)}]{Kang:2008ea}%
  \BibitemOpen
  \bibfield  {author} {\bibinfo {author} {\bibfnamefont {J.}~\bibnamefont
  {Kang}}\ and\ \bibinfo {author} {\bibfnamefont {M.~A.}\ \bibnamefont
  {Luty}},\ }\href {\doibase 10.1088/1126-6708/2009/11/065} {\bibfield
  {journal} {\bibinfo  {journal} {JHEP}\ }\textbf {\bibinfo {volume} {0911}},\
  \bibinfo {pages} {065} (\bibinfo {year} {2009})},\ \Eprint
  {http://arxiv.org/abs/0805.4642} {arXiv:0805.4642 [hep-ph]} \BibitemShut
  {NoStop}%
\bibitem [{\citenamefont {Burdman}\ \emph {et~al.}(2008)\citenamefont
  {Burdman}, \citenamefont {Chacko}, \citenamefont {Goh}, \citenamefont
  {Harnik},\ and\ \citenamefont {Krenke}}]{Burdman:2008ek}%
  \BibitemOpen
  \bibfield  {author} {\bibinfo {author} {\bibfnamefont {G.}~\bibnamefont
  {Burdman}}, \bibinfo {author} {\bibfnamefont {Z.}~\bibnamefont {Chacko}},
  \bibinfo {author} {\bibfnamefont {H.-S.}\ \bibnamefont {Goh}}, \bibinfo
  {author} {\bibfnamefont {R.}~\bibnamefont {Harnik}}, \ and\ \bibinfo {author}
  {\bibfnamefont {C.~A.}\ \bibnamefont {Krenke}},\ }\href {\doibase
  10.1103/PhysRevD.78.075028} {\bibfield  {journal} {\bibinfo  {journal}
  {Phys.Rev.}\ }\textbf {\bibinfo {volume} {D78}},\ \bibinfo {pages} {075028}
  (\bibinfo {year} {2008})},\ \Eprint {http://arxiv.org/abs/0805.4667}
  {arXiv:0805.4667 [hep-ph]} \BibitemShut {NoStop}%
\bibitem [{\citenamefont {Harnik}\ and\ \citenamefont
  {Wizansky}(2009)}]{Harnik:2008ax}%
  \BibitemOpen
  \bibfield  {author} {\bibinfo {author} {\bibfnamefont {R.}~\bibnamefont
  {Harnik}}\ and\ \bibinfo {author} {\bibfnamefont {T.}~\bibnamefont
  {Wizansky}},\ }\href {\doibase 10.1103/PhysRevD.80.075015} {\bibfield
  {journal} {\bibinfo  {journal} {Phys.Rev.}\ }\textbf {\bibinfo {volume}
  {D80}},\ \bibinfo {pages} {075015} (\bibinfo {year} {2009})},\ \Eprint
  {http://arxiv.org/abs/0810.3948} {arXiv:0810.3948 [hep-ph]} \BibitemShut
  {NoStop}%
\bibitem [{\citenamefont {Djouadi}(2008)}]{Djouadi20081}%
  \BibitemOpen
  \bibfield  {author} {\bibinfo {author} {\bibfnamefont {A.}~\bibnamefont
  {Djouadi}},\ }\href {\doibase
  http://dx.doi.org/10.1016/j.physrep.2007.10.004} {\bibfield  {journal}
  {\bibinfo  {journal} {Physics Reports}\ }\textbf {\bibinfo {volume} {457}},\
  \bibinfo {pages} {1 } (\bibinfo {year} {2008})},\ \Eprint
  {http://arxiv.org/abs/hep-ph/05031720} {arXiv:hep-ph/05031720 [hep-ph]}
  \BibitemShut {NoStop}%
\bibitem [{\citenamefont {Heinemeier}\ \emph {et~al.}(2013)\citenamefont
  {Heinemeier}, \citenamefont {Mariotti}, \citenamefont {Passarino},\ and\
  \citenamefont {Tanaka}}]{HiggsProp2013}%
  \BibitemOpen
  \bibfield  {author} {\bibinfo {author} {\bibfnamefont {S.}~\bibnamefont
  {Heinemeier}}, \bibinfo {author} {\bibfnamefont {C.}~\bibnamefont
  {Mariotti}}, \bibinfo {author} {\bibfnamefont {G.}~\bibnamefont {Passarino}},
  \ and\ \bibinfo {author} {\bibfnamefont {R.}~\bibnamefont {Tanaka}} (\bibinfo
  {collaboration} {LHC Higgs Cross Section Working Group}),\ }\href@noop {}
  {\bibfield  {journal} {\bibinfo  {journal} {CERN–2013–004}\ } (\bibinfo
  {year} {2013})},\ \Eprint {http://arxiv.org/abs/arXiv:1307.1347 [hep-ph]}
  {arXiv:1307.1347 [hep-ph]} \BibitemShut {NoStop}%
\bibitem [{\citenamefont {Falkowski}\ \emph {et~al.}(2013)\citenamefont
  {Falkowski}, \citenamefont {Riva},\ and\ \citenamefont
  {Urbano}}]{Falkowski:2013dza}%
  \BibitemOpen
  \bibfield  {author} {\bibinfo {author} {\bibfnamefont {A.}~\bibnamefont
  {Falkowski}}, \bibinfo {author} {\bibfnamefont {F.}~\bibnamefont {Riva}}, \
  and\ \bibinfo {author} {\bibfnamefont {A.}~\bibnamefont {Urbano}},\ }\href
  {\doibase 10.1007/JHEP11(2013)111} {\bibfield  {journal} {\bibinfo  {journal}
  {JHEP}\ }\textbf {\bibinfo {volume} {1311}},\ \bibinfo {pages} {111}
  (\bibinfo {year} {2013})},\ \Eprint {http://arxiv.org/abs/1303.1812}
  {arXiv:1303.1812 [hep-ph]} \BibitemShut {NoStop}%
\bibitem [{\citenamefont {Agashe}\ \emph {et~al.}(2005)\citenamefont {Agashe},
  \citenamefont {Contino},\ and\ \citenamefont {Pomarol}}]{Agashe:2004rs}%
  \BibitemOpen
  \bibfield  {author} {\bibinfo {author} {\bibfnamefont {K.}~\bibnamefont
  {Agashe}}, \bibinfo {author} {\bibfnamefont {R.}~\bibnamefont {Contino}}, \
  and\ \bibinfo {author} {\bibfnamefont {A.}~\bibnamefont {Pomarol}},\ }\href
  {\doibase 10.1016/j.nuclphysb.2005.04.035} {\bibfield  {journal} {\bibinfo
  {journal} {Nucl.Phys.}\ }\textbf {\bibinfo {volume} {B719}},\ \bibinfo
  {pages} {165} (\bibinfo {year} {2005})},\ \Eprint
  {http://arxiv.org/abs/hep-ph/0412089} {arXiv:hep-ph/0412089 [hep-ph]}
  \BibitemShut {NoStop}%
\bibitem [{\citenamefont {Dawson}\ \emph {et~al.}(2013)\citenamefont {Dawson},
  \citenamefont {Gritsan}, \citenamefont {Logan}, \citenamefont {Qian},
  \citenamefont {Tully} \emph {et~al.}}]{Dawson:2013bba}%
  \BibitemOpen
  \bibfield  {author} {\bibinfo {author} {\bibfnamefont {S.}~\bibnamefont
  {Dawson}}, \bibinfo {author} {\bibfnamefont {A.}~\bibnamefont {Gritsan}},
  \bibinfo {author} {\bibfnamefont {H.}~\bibnamefont {Logan}}, \bibinfo
  {author} {\bibfnamefont {J.}~\bibnamefont {Qian}}, \bibinfo {author}
  {\bibfnamefont {C.}~\bibnamefont {Tully}},  \emph {et~al.},\ }\href@noop {}
  {\  (\bibinfo {year} {2013})},\ \Eprint {http://arxiv.org/abs/1310.8361}
  {arXiv:1310.8361 [hep-ex]} \BibitemShut {NoStop}%
\bibitem [{\citenamefont {Craig}\ and\ \citenamefont
  {Lou}(2014)}]{Craig:2014fka}%
  \BibitemOpen
  \bibfield  {author} {\bibinfo {author} {\bibfnamefont {N.}~\bibnamefont
  {Craig}}\ and\ \bibinfo {author} {\bibfnamefont {H.~K.}\ \bibnamefont
  {Lou}},\ }\href@noop {} {\  (\bibinfo {year} {2014})},\ \Eprint
  {http://arxiv.org/abs/1406.4880} {arXiv:1406.4880 [hep-ph]} \BibitemShut
  {NoStop}%
\bibitem [{\citenamefont {Haber}(1993)}]{HaberSusy93}%
  \BibitemOpen
  \bibfield  {author} {\bibinfo {author} {\bibfnamefont {H.~E.}\ \bibnamefont
  {Haber}},\ }\href@noop {} {\  (\bibinfo {year} {1993})},\ \Eprint
  {http://arxiv.org/abs/hep-ph/9306207} {arXiv:hep-ph/9306207 [hep-ph]}
  \BibitemShut {NoStop}%
\bibitem [{\citenamefont {Batra}\ \emph {et~al.}(2004)\citenamefont {Batra},
  \citenamefont {Delgado}, \citenamefont {Kaplan},\ and\ \citenamefont
  {Tait}}]{Batra:2003nj}%
  \BibitemOpen
  \bibfield  {author} {\bibinfo {author} {\bibfnamefont {P.}~\bibnamefont
  {Batra}}, \bibinfo {author} {\bibfnamefont {A.}~\bibnamefont {Delgado}},
  \bibinfo {author} {\bibfnamefont {D.~E.}\ \bibnamefont {Kaplan}}, \ and\
  \bibinfo {author} {\bibfnamefont {T.~M.}\ \bibnamefont {Tait}},\ }\href
  {\doibase 10.1088/1126-6708/2004/02/043} {\bibfield  {journal} {\bibinfo
  {journal} {JHEP}\ }\textbf {\bibinfo {volume} {0402}},\ \bibinfo {pages}
  {043} (\bibinfo {year} {2004})},\ \Eprint
  {http://arxiv.org/abs/hep-ph/0309149} {arXiv:hep-ph/0309149 [hep-ph]}
  \BibitemShut {NoStop}%
\bibitem [{\citenamefont {Olive}\ \emph {et~al.}(2014)\citenamefont {Olive}
  \emph {et~al.}}]{Agashe:2014kda}%
  \BibitemOpen
  \bibfield  {author} {\bibinfo {author} {\bibfnamefont {K.}~\bibnamefont
  {Olive}} \emph {et~al.} (\bibinfo {collaboration} {Particle Data Group}),\
  }\href {\doibase 10.1088/1674-1137/38/9/090001} {\bibfield  {journal}
  {\bibinfo  {journal} {Chin.Phys.}\ }\textbf {\bibinfo {volume} {C38}},\
  \bibinfo {pages} {090001} (\bibinfo {year} {2014})}\BibitemShut {NoStop}%
\bibitem [{\citenamefont {Fan}\ and\ \citenamefont
  {Reece}(2014)}]{Fan:2014txa}%
  \BibitemOpen
  \bibfield  {author} {\bibinfo {author} {\bibfnamefont {J.}~\bibnamefont
  {Fan}}\ and\ \bibinfo {author} {\bibfnamefont {M.}~\bibnamefont {Reece}},\
  }\href {\doibase 10.1007/JHEP06(2014)031} {\bibfield  {journal} {\bibinfo
  {journal} {JHEP}\ }\textbf {\bibinfo {volume} {1406}},\ \bibinfo {pages}
  {031} (\bibinfo {year} {2014})},\ \Eprint {http://arxiv.org/abs/1401.7671}
  {arXiv:1401.7671 [hep-ph]} \BibitemShut {NoStop}%
\bibitem [{\citenamefont {Aad}\ \emph {et~al.}(2014{\natexlab{c}})\citenamefont
  {Aad} \emph {et~al.}}]{Aad:2014fha}%
  \BibitemOpen
  \bibfield  {author} {\bibinfo {author} {\bibfnamefont {G.}~\bibnamefont
  {Aad}} \emph {et~al.} (\bibinfo {collaboration} {ATLAS Collaboration}),\
  }\href {\doibase 10.1016/j.physletb.2014.10.002} {\bibfield  {journal}
  {\bibinfo  {journal} {Phys.Lett.}\ }\textbf {\bibinfo {volume} {B738}},\
  \bibinfo {pages} {428} (\bibinfo {year} {2014}{\natexlab{c}})},\ \Eprint
  {http://arxiv.org/abs/1407.8150} {arXiv:1407.8150 [hep-ex]} \BibitemShut
  {NoStop}%
\bibitem [{\citenamefont {Alwall}\ \emph {et~al.}(2011)\citenamefont {Alwall},
  \citenamefont {Herquet}, \citenamefont {Maltoni}, \citenamefont {Mattelaer},\
  and\ \citenamefont {Stelzer}}]{Alwall:2011uj}%
  \BibitemOpen
  \bibfield  {author} {\bibinfo {author} {\bibfnamefont {J.}~\bibnamefont
  {Alwall}}, \bibinfo {author} {\bibfnamefont {M.}~\bibnamefont {Herquet}},
  \bibinfo {author} {\bibfnamefont {F.}~\bibnamefont {Maltoni}}, \bibinfo
  {author} {\bibfnamefont {O.}~\bibnamefont {Mattelaer}}, \ and\ \bibinfo
  {author} {\bibfnamefont {T.}~\bibnamefont {Stelzer}},\ }\href {\doibase
  10.1007/JHEP06(2011)128} {\bibfield  {journal} {\bibinfo  {journal} {JHEP}\
  }\textbf {\bibinfo {volume} {1106}},\ \bibinfo {pages} {128} (\bibinfo {year}
  {2011})},\ \Eprint {http://arxiv.org/abs/1106.0522} {arXiv:1106.0522
  [hep-ph]} \BibitemShut {NoStop}%
\end{thebibliography}%

\end{document}